\documentclass[onecolumn,showpacs,preprintnumbers,amsmath,amssymb]{revtex4}
\usepackage{graphicx}
\usepackage{dcolumn}
\usepackage{bm}
\usepackage{color}
\begin{document}
%
%%%%%%%%%%%%%%%%%%%%%%%%%%%%
\newcommand{\kvec}{\mbox{{\scriptsize {\bf k}}}}
\newcommand{\lvec}{\mbox{{\scriptsize {\bf l}}}}
\newcommand{\qvec}{\mbox{{\scriptsize {\bf q}}}}
%%%%%%%%%%%%%%%%%%%%%%%%%%%%
%
%%%%%%%%%%%%%%%%%%%%%%%%%%%%
\def\eq#1{\hspace{1mm}(\ref{#1})}
\def\fig#1{\hspace{1mm}\ref{#1}}
\def\tab#1{\hspace{1mm}\ref{#1}}
%%%%%%%%%%%%%%%%%%%%%%%%%%%%
%%%%%%%%%%%%%%%%%%%%%%%%%%%%
\title{Anisotropy of the gap parameter in the hole-doped cuprates}
\author{R. Szcz{\c{e}}{\'s}niak} \email{szczesni@wip.pcz.pl}
\author{A.P. Durajski} \email{adurajski@wip.pcz.pl}
\affiliation{Institute of Physics, Cz{\c{e}}stochowa University of Technology, Ave. Armii Krajowej 19, 42-200 Cz{\c{e}}stochowa, Poland}
\date{\today}
%%%%%%%%%%%%%%%
\begin{abstract}
The structure of the gap parameter ($\Delta_{\kvec}$) for the hole-doped cuprates has been studied. The obtained results indicate that the antinodal part of $\Delta_{\kvec}$ is very weakly temperature dependent and above the critical temperature ($T_{C}$), it extends into the anomalous normal state to the pseudogap temperature. On the other hand, the values of $\Delta_{\kvec}$, which are close to the nodal part, are strongly temperature dependent. The model has been tested for the ${\rm YBa_{2}Cu_{3}O_{7-\delta}}$ superconductor. It has been shown that the theoretical results agree with the experimental data. 
\end{abstract}
%%%%%%%%%%%%%%%
\maketitle

The gap parameter ($\Delta_{\kvec}$) is strongly anisotropic in the cuprates \cite{Bednorz1986A}, \cite{Bednorz1988A}. In the case of the hole-doped superconductors, the $d$-wave symmetry is dominant \cite{Dagotto1994A}, \cite{Harlingen1995A}, \cite{Tsuei2000A}. In particular, for the optimally doped ${\rm YBa_{2}Cu_{3}O_{7-\delta}}$, the experimental data suggests $83\%$ $d$-wave anisotropy \cite{Smilde2005A}. 

The issue of the wave symmetry for the electron-doped compounds is still unclear. Some experiments favor the $s$-wave symmetry \cite{Armitage2010A}, 
others support the $d$-wave anisotropy (penetration depth \cite{Kokales2000A}, \cite{Prozorov2000A}, tricrystal \cite{Tsuei2000B}, photoemission \cite{Armitage2001A}, \cite{Sato2001A}, Raman scattering \cite{Blumberg2002A}, and point contact tunneling \cite{Biswas2002A}). Additionally, the point contact tunneling data \cite{Biswas2002A} and the penetration depth experiment \cite{Skinta2002A} suggest the energy gap crossover from the $d$- to $s$-wave symmetry, as the doping changes from the underdoped to overdoped region.

The gap parameter for the hole-doped cuprates can be most precisely measured by using the ARPES method 
\cite{Damascelli2003A}, \cite{Yoshida2012A}, \cite{Hashimoto2012A}, \cite{Hashimoto2014A}, \cite{Vishik2014A}. So far, the obtained results have been interpreted in the framework of the two different approaches. 

In the first case, the difference between the doping and temperature dependence of the gap in the nodal and antinodal region suggests that the pseudogap and the superconducting gap are independent \cite{Tanaka2006A}, \cite{Lee2007A}. Additional support comes from the strong deviation from the standard $d$-wave form of the energy gap in the underdoped region. The above fact is interpreted as composition of the $d$-wave superconducting gap and the remnant pseudogap \cite{Kondo2007A}, \cite{Terashima2007A}. 

In the second case, the pseudogap is considered as the precursor of the superconducting gap \cite{Kanigel2007A}, \cite{Shi2008A}. The $d$-wave symmetry deviation in the underdoped region is connected with the existence of the high-harmonic pairing terms \cite{Mesot1999A}.

In the presented paper, we have studied the anisotropy of the gap parameter for the hole-doped superconductors in the framework of the recently introduced theory \cite{Szczesniak2012D}, \cite{Szczesniak2012E}. 

Our main purpose was to derivation the thermodynamic equation for the anomalous thermal average, which determines the structure of the gap parameter. Next, the temperature dependence of the nodal and the antinodal part of the energy gap has been calculated. We have assumed that the theory should be simple enough as far as possible. However, the good agreement between the theoretical predictions and the experimental results has been also required.

The model is based on three postulates:
(i) {\it In the superconductivity domain of the cuprates the fundamental role is played by the electrons on the $CuO_{2}$ planes.}
(ii) {\it The conventional electron-phonon {\rm(EPH)} interaction exists in the cuprates, which does not have to be strong.}
(iii) {\it Strong electronic correlations exist in the cuprates, but the electron-electron scattering in the superconductivity domain is inseparably connected with absorption or emission of the vibrational quanta.} 

The first postulate emphasizes the importance of the quasi two-dimensionality of the system. The second one refers to the classical pairing mechanism given by Fr{\"o}hlich \cite{Frohlich1950A}, \cite{Frohlich1954A}. The third postulate states that the strong electron correlations in the cuprates are inseparably coupled with the phonon subsystem. The first two postulates define the van Hove scenario \cite{Szczesniak2001A}, \cite{Szczesniak2006B}. The third postulate requires further discussion because it is far more subtle. In particular, it should be noted that the postulated electron correlations generalize the Hubbard approach \cite{Hubbard1963A}; the classical two-body interaction is replaced with the three-body interaction (the electron-electron-phonon (EEPH)). It should be pointed out very clearly that the third postulate does not require the additional phonon channel to be as strong as the electron channel.
 
It has been shown, that for strong EEPH coupling and $T < T_{C}$, the average gap parameter ($\Delta_{tot}$) is very weakly temperature dependent, and up to the critical temperature extends into the anomalous normal state: (i) in the $s$-wave case to the Nernst temperature, (ii) for the $d$-wave symmetry to the pseudogap temperature. We boldly underline that the discussed model explains well the experimental dependence of the ratio $R_{1}\equiv 2\Delta_{tot}^{\left(0\right)}/k_{B}T_{C}$ on doping for the reported superconductors in the terms of the few fundamental parameters calculated by using the {\it ab initio} methods \cite{Szczesniak2012D}. The symbol $\Delta_{tot}^{\left(0\right)}$ denotes the average gap parameter close to temperature of the zero Kelvin.

\vspace*{0.25cm}

%%%%%%%%%%%%%%%%%%%%%%%%%%%%%%%%%%%%%%%%%%%%%%%%%%%%%%%%%%%%%%%%%%%%%%%%%%%%%%%%%%%%%%%%%%%%%%%%%%%%%%%%%%%%%%%%%%%%%%%%%%%%%%%%%%elektro-elektron interakcja

Note the fact that the pairing mechanism can also be connected with the pure electron-electron interaction. This approach is based on the Hubbard-like or related models ({\it e.g.} $t-J$ model) \cite{Hubbard1963A}, \cite{Emery1987A}, \cite{Littlewood1987A}, \cite{Anderson1987A}, \cite{Millis1990A}, \cite{Monthoux1992A}, \cite{Lee2006A}, \cite{Chao1977A}.

In particular, the theoretical results suggest that the one-band Hubbard operator reproduces well the spectra of the three-band Emery Hamiltonian for electrons in the copper oxide planes \cite{Dagotto1994A}, \cite{Emery1987A}, \cite{Littlewood1987A}. Additionally, for the half-filled electron band and large on-site Coulomb interaction, the Hubbard Hamiltonian reduces to the Heisenbeg operator, which describes well the spin dynamics of the underdoped cuprates (antiferromagnetism) \cite{Dagotto1994A}.

Unfortunately, the one-band Hubbard model gives no obvious evidence for superconductivity with the high critical temperature \cite{Imada1989A}. On the other hand, the high-$T_{C}$ superconductivity exists in the attractive Hubbard model for the same value of the on-site Coulomb interaction. Probably also the Emery and  $t-J$ model do not superconduct at temperatures characteristic for the high-$T_{C}$ superconductors \cite{Dagotto1994A}, \cite{Pryadko2004A}.

We also underline that the Hubbard approach ignores completely many experimental data which have been taken as evidence for the electron-phonon interaction in the cuprates \cite{Damascelli2003A}, \cite{Vedeneev1995A},\cite{Hofer2000A},\cite{Kulic2000A},\cite{Cuk2005A}.
However, it should be clearly emphasized the importance of the models of strongly correlated fermions that can provide single description of both magnetism and the paired state.

%%%%%%%%%%%%%%%%%%%%%%%%%%%%%%%%%%%%%%%%%%%%%%%%%%%%%%%%%%%%%%%%%%%%%%%%%%%%%%%%%%%%%%%%%%%%%%%%%%%%%%%%%%%%%%%%%%%%%%%%%%%%%%%%%%%%%%%%%%%%%%%%%%%%%%%%%%%%%        

\vspace*{0.25cm}

%%%%%%%%%%%%%%%%%%%%%%%%%%%%%%%%%%%%%%%%%%%%%%%%%%%%%%%%%%%%%%%%%%%%%%%%%%%%%%%%%%%%%%%%%%%%%%%%%%%%%%%%%%%%%%%%%%%%%%%%%%%%%%%%%%%%%%%%%%(Hamiltonian step I)

In the presented approach, the Hamiltonian takes the form \cite{Szczesniak2012D}:
\begin{equation}
\label{r-1}
H\equiv H^{\left(0\right)}+H^{\left(1\right)}+H^{\left(2\right)},
\end{equation}
where $H^{\left(0\right)}$ describes the non-interacting electrons and phonons:
\begin{equation}
\label{r-2}
H^{\left(0\right)}\equiv\sum_{\kvec\sigma }\varepsilon _{\kvec}c_{\kvec\sigma
}^{\dagger}c_{\kvec\sigma }+\sum_{\qvec}\omega _{\qvec}b_{\qvec}^{\dagger}b_{\qvec}.
\end{equation}

The band energy for the two-dimensional square lattice can be written as: $\varepsilon _{\kvec}=-t\gamma\left({\bf k}\right)$, where $t$ denotes the nearest-neighbour hopping integral and $\gamma\left({\bf k}\right)\equiv 2\left[\cos\left(k_{x}\right)+\cos\left(k_{y}\right)\right]$. 
We notice that in the cuprates the non-zero value possesses also the second-neighbour hopping integral ($t'$). However, the value of $t'$ is much lower than the value of the nearest-neighbour hopping integral \cite{Zhang1988A}, \cite{Eskes1988A}, \cite{Hybertsen1988A}, \cite{Rice1991A}. For this reason, in the first approximation, we have neglected $t'$.
The symbols $c^{\dagger}_{\kvec\sigma}$ and $c_{\kvec\sigma}$ are the creation and annihilation operators for the electron with momentum ${\bf k}$ and spin $\sigma$. The function $\omega_{\qvec}$ models the energy of the phonon with the wave number ${\bf q}$. The operators $b^{\dagger}_{\qvec}$, $b_{\qvec}$ are the phonon creation and annihilation operators, respectively.

The electron-phonon (EPH) and electron-electron-phonon (EEPH) terms are given by:
\begin{equation}
\label{r-3}
H^{\left(1\right)}\equiv\sum_{\kvec\qvec\sigma }g^{\left(1\right)}_{\kvec}\left({\bf q}\right)
c_{\kvec+\qvec\sigma}^{\dagger}c_{\kvec\sigma}\varphi_{\qvec},
\end{equation}
and
\begin{equation}
\label{r-4}
H^{\left(2\right)}\equiv\sum_{\kvec\kvec^{'}\qvec\lvec\sigma}
g^{\left(2\right)}_{\kvec,\kvec^{'}}\left({\bf q},{\bf l}\right)
c_{\kvec-\lvec\sigma }^{\dagger}c_{\kvec\sigma}
c_{\kvec^{'}+\lvec+\qvec-\sigma}^{\dagger}c_{\kvec^{'}-\sigma}\varphi_{\qvec},
\end{equation}
where: $\varphi_{\qvec}\equiv b_{-\qvec}^{\dagger}+b_{\qvec}$. The symbol $g^{\left(1\right)}_{\kvec}\left({\bf q}\right)\simeq g^{\left(1\right)}$ denotes EPH coupling \cite{ Frohlich1950A}, \cite{ Frohlich1954A}, and 
$g^{\left(2\right)}_{\kvec,\kvec^{'}}\left({\bf q},{\bf l}\right)\simeq  g^{\left(2\right)}$ models the EEPH interaction.

%%%%%%%%%%%%%%%%%%%%%%%%%%%%%%%%%%%%%%%%%%%%%%%%%%%%%%%%%%%%%%%%%%%%%%%%%%%%%%%%%%%%%%%%%%%%%%%%%%%%%%%%%%%%%%%%%%%%%%%%%%%%%%%%%%%%%%%%%(Hamiltonian step II)

In order to simplify the Hamiltonian \eq{r-1}, the phonon degrees of freedom have been eliminated by using the unitary transformation. 
The effective Hamiltonian can be written as \cite{Szczesniak2012D}:
\begin{equation}
\label{r-5}
H^{'}\simeq H^{'\left(0\right)}+H^{'\left(1\right)}+H^{'\left(2\right)},
\end{equation}
where:
\begin{equation}
\label{r-6}
H^{'\left(0\right)}\equiv\sum_{\kvec\sigma }\varepsilon _{\kvec}c_{\kvec\sigma}^{\dagger}c_{\kvec\sigma },
\end{equation}
\begin{equation}
\label{r-7}
H^{'\left(1\right)}\equiv- \frac{V}{2N_{0}}\sum^{\omega_{0}}_{\kvec\qvec\sigma}
c^{\dagger}_{\kvec+\qvec-\sigma}c^{\dagger}_{-\kvec-\qvec\sigma}
c_{-\kvec\sigma}c_{\kvec-\sigma},
\end{equation}
and
\begin{equation}
\label{r-8}
H^{'\left(2\right)}\equiv-\frac{U}{4N_{0}^{3}}
\sum^{\omega_{0}}_{\kvec\kvec^{'}\qvec\lvec\sigma}
c_{\kvec-\lvec\sigma }^{\dagger}c_{\kvec\sigma}
h_{\kvec^{'}\lvec\qvec\sigma}
c_{-\kvec+\lvec-\sigma }^{\dagger}c_{-\kvec-\sigma},
\end{equation}
The operator $h_{\kvec^{'}\lvec\qvec\sigma}$ has been defined by the expression: 
$h_{\kvec^{'}\lvec\qvec\sigma}\equiv c_{\kvec^{'}+\lvec+\qvec-\sigma }^{\dagger}
c_{-\kvec^{'}-\lvec-\qvec\sigma }^{\dagger}c_{-\kvec^{'}\sigma}c_{\kvec^{'}-\sigma}$.

The symbols $V$ and $U$ represent the attractive parts of the pairing potentials:
\begin{equation}
\label{r-9}
V_{\kvec\qvec}\equiv\frac{\omega_{0}|g^{\left(1\right)}|^{2}}
{\left(\varepsilon_{\kvec}-\varepsilon_{\kvec+\qvec}\right)^{2}-\omega^{2}_{0}}\rightarrow - \frac{V}{2N_{0}},
\end{equation}
and
\begin{equation}
\label{r-10}
U_{\kvec\kvec^{'}\qvec\lvec}\equiv
\frac{\omega_{0}|g^{\left(2\right)}|^{2}}
{\left(\varepsilon_{\kvec}-\varepsilon_{\kvec-\lvec}+\varepsilon_{\kvec^{'}}-
\varepsilon_{\kvec^{'}+\lvec+\qvec}\right)^{2}-\omega^{2}_{0}}\rightarrow -\frac{U}{4N_{0}^{3}}.
\end{equation}

The parameter $N_{0}$ is the normalization factor: $N_{0}\equiv 1/\sum^{\omega_{0}}_{\kvec}$; the symbol $\sum^{\omega_{0}}_{\kvec}$ denotes the sum over the states for which: $|\varepsilon_{\kvec}|\leq\omega_{0}$. The quantity $\omega_{0}$ represents the characteristic phonon frequency ($\omega_{0}$ is of the order of Debye frequency).

%%%%%%%%%%%%%%%%%%%%%%%%%%%%%%%%%%%%%%%%%%%%%%%%%%%%%%%%%%%%%%%%%%%%%%%%%%%%%%%%%%%%%%%%%%%%%%%%%%%%%%%%%%%%%%%%%%%%%%%%%%%%%%%%%%%%%%%%(Hamiltonian step III)

The Hamiltonian \eq{r-8} possesses still complicated form (this term has eight fermion operators). Thus, we have simplified Eq. \eq{r-8} by using the procedure presented in \cite{Szczesniak2012D}. In particular, on the basis of the expression: 
$AB\simeq \left<A\right>B+A\left<B\right>-\left<A\right>\left<B\right>$, we can written:
\begin{eqnarray}
\label{r-11}
h_{\kvec^{'}\lvec\qvec\sigma}&\simeq& 
\phi_{\kvec^{'}\sigma}
c_{\kvec^{'}+\lvec+\qvec-\sigma }^{\dagger}
c_{-\kvec^{'}-\lvec-\qvec\sigma }^{\dagger}\\ \nonumber
&+&
\phi^{\star}_{\kvec^{'}+\lvec+\qvec\sigma}
c_{-\kvec^{'}\sigma}c_{\kvec^{'}-\sigma}\\ \nonumber
&-&
\phi_{\kvec^{'}\sigma}\phi^{\star}_{\kvec^{'}+\lvec+\qvec\sigma}.
\end{eqnarray}
The anomalous thermal average is defined as: $\phi_{\kvec\sigma}\equiv\left<c_{-\kvec\sigma}c_{\kvec-\sigma}\right>$.
Now, we insert Eq. \eq{r-11} into Eq. \eq{r-8}:
\begin{eqnarray}
\label{r-12}
H^{'\left(2\right)}&\simeq&
\frac{U}{4N_{0}^{3}}\sum^{\omega_{0}}_{\kvec\kvec^{'}\qvec\lvec\sigma}\phi^{\star}_{\kvec^{'}+\lvec+\qvec\sigma}c_{\kvec^{'}-\sigma}
h^{'}_{\kvec\lvec\sigma}c_{-\kvec^{'}\sigma}\\ \nonumber
&-&
\frac{U}{4N_{0}^{3}}\sum^{\omega_{0}}_{\kvec\kvec^{'}\qvec\lvec\sigma}\phi_{\kvec^{'}\sigma}c^{\dagger}_{\kvec^{'}+\lvec+\qvec-\sigma}
h^{'}_{\kvec\lvec\sigma}c^{\dagger}_{-\kvec^{'}-\lvec-\qvec\sigma}\\ \nonumber
&+&
\frac{U}{4N_{0}^{3}}\sum^{\omega_{0}}_{\kvec\kvec^{'}\qvec\lvec\sigma}\phi^{\star}_{\kvec^{'}+\lvec+\qvec\sigma}\phi_{\kvec^{'}\sigma}
h^{'}_{\kvec\lvec\sigma},
\end{eqnarray}
where: 
$h^{'}_{\kvec\lvec\sigma}\equiv c_{-\kvec+\lvec-\sigma }^{\dagger}c_{\kvec-\lvec\sigma }^{\dagger}c_{\kvec\sigma}c_{-\kvec-\sigma}$.

In the considered case, we simplify the operator \eq{r-12} again:
\begin{eqnarray}
\label{r-13}
h^{'}_{\kvec\lvec\sigma}&\simeq& 
\phi_{-\kvec\sigma}c_{-\kvec+\lvec-\sigma }^{\dagger}c_{\kvec-\lvec\sigma }^{\dagger}\\ \nonumber
&+&
\phi^{\star}_{-\kvec+\lvec\sigma}c_{\kvec\sigma}c_{-\kvec-\sigma}\\ \nonumber
&-&
\phi_{-\kvec\sigma}\phi^{\star}_{-\kvec+\lvec\sigma}.
\end{eqnarray}

Finally, we have:
\begin{eqnarray}
\label{r-14}
H^{'\left(2\right)}&\simeq& 
-\frac{U}{2N_{0}^{3}}\sum^{\omega_{0}}_{\kvec\kvec^{'}\qvec\lvec\sigma}
\phi_{\kvec\sigma}\phi^{\star}_{\kvec^{'}+\lvec\sigma} \\ \nonumber
&\times&
c_{-\kvec^{'}\sigma}c_{\kvec^{'}-\sigma}
c_{\kvec+\lvec+\qvec-\sigma}^{\dagger}c_{-\kvec-\lvec-\qvec\sigma}^{\dagger}.
\end{eqnarray}
%

%%%%%%%%%%%%%%%%%%%%%%%%%%%%%%%%%%%%%%%%%%%%%%%%%%%%%%%%%%%%%%%%%%%%%%%%%%%%%%%%%%%%%%%%%%%%%%%%%%%%%%%%%%%%%%%%%%%%%%%%%%%%%%%%%%%%%%%%%(Hamiltonian step IV)

On the basis of the Eqs. \eq{r-7} and \eq{r-14}, it is possible to deduce the Hamiltonian, which describe the $d$-wave superconducting state \cite{Newns}.

In particular, with help of the transformation: $c_{\kvec\sigma}=\frac{1}{\sqrt{N}}\sum_{j}e^{-i\kvec {\bf R}_{j}}c_{j\sigma}$, we have rewritten the operators \eq{r-7} and \eq{r-14} to the Wannier representation, where we have focused ourselves to on-site and the nearest neighbour pairing. Next, the symmetry has been limited to the dominating $d$-wave symmetry. Finally, we have separated the momentums in the considered expressions.
%
%%%%%%%%%%%%%%%%%%%%%%%%%%%%%%%%%%%%%%%%%%%%%%%%%%%%%%%%%%%%%%%%%%%%%%%%%%%%%%%%%%%%%%%%%%%%%%%%%%%%%%%%%%%%%%%%%%%%%%%%%%%%%%%%%%%%%%%%%%(Hamiltonian step V)
%
The final form of the Hamiltonian can be written as:
\begin{equation}
\label{r1}
H^{''}\equiv H^{'\left(0\right)}+H^{''\left(1\right)}+H^{''\left(2\right)},
\end{equation}
\begin{equation}
\label{r3}
H^{''\left(1\right)}\equiv-\sum^{\omega_{0}}_{\kvec_{1}\kvec_{2}\sigma}\frac{V_{\kvec_{1}\kvec_{2}}}{2N_{0}}
c^{\dagger}_{\kvec_{1}-\sigma}c^{\dagger}_{-\kvec_{1}\sigma}c_{-\kvec_{2}\sigma}c_{\kvec_{2}-\sigma},
\end{equation}
and
\begin{equation}
\label{r4}
H^{''\left(2\right)}\equiv-\sum^{\omega_{0}}_{\kvec_{1}\sim\kvec_{4}\sigma}
\frac{U_{\kvec_{1}\sim\kvec_{4}}}{2N^{3}_{0}}\phi_{\kvec_{1}\sigma}\phi^{\star}_{\kvec_{2}\sigma}
c_{-\kvec_{3}\sigma }c_{\kvec_{3}-\sigma}c^{\dagger}_{\kvec_{4}-\sigma }c^{\dagger}_{-\kvec_{4}\sigma}.
\end{equation}
The functions $V_{\kvec_{1}\kvec_{2}}$ and $U_{\kvec_{1}\sim\kvec_{4}}$ represent the pairing potentials with the $d$-wave symmetry:
\begin{equation}
\label{r5}
V_{\kvec_{1}\kvec_{2}}\equiv V\eta\left({\bf k_{1}}\right)\eta\left({\bf k_{2}}\right),
\end{equation}
and
\begin{equation}
\label{r6}
U_{\kvec_{1}\sim\kvec_{4}}\equiv U\eta\left({\bf k_{1}}\right)\eta\left({\bf k_{2}}\right)\eta\left({\bf k_{3}}\right)\eta\left({\bf k_{4}}\right),
\end{equation}
where: $\eta\left({\bf k}\right)\equiv 2\left[\cos\left(k_{x}\right)-\cos\left(k_{y}\right)\right]$. 
\begin{figure}
\centering
\includegraphics[scale=0.19]{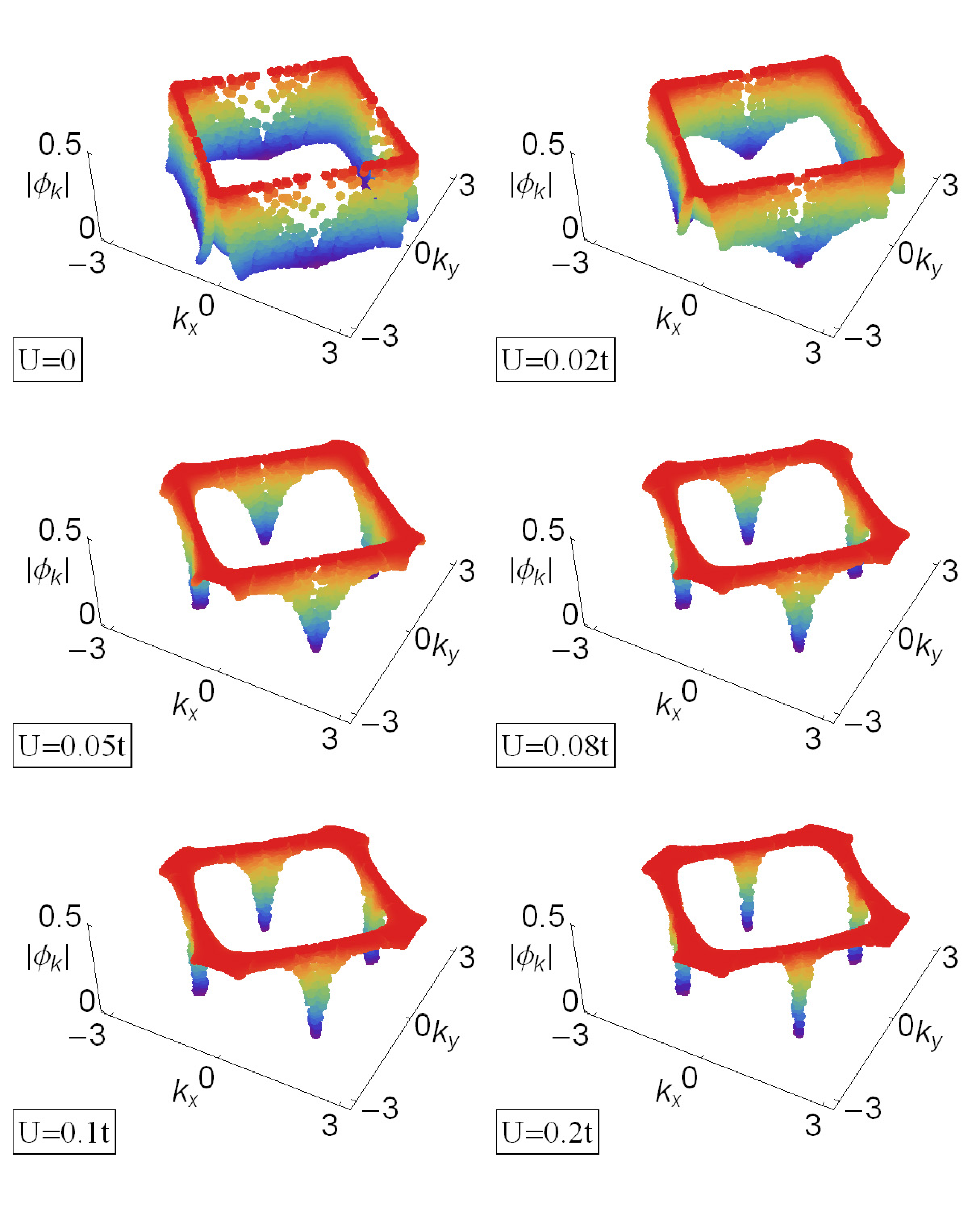}
\caption{The amplitude of the anomalous thermal average close to the Fermi energy for selected values of the potential $U$. We have assumed: $V=0.02t$ and $\omega_{0}=0.3t$.}
\label{f1}
\end{figure}
\begin{figure}
\centering
\includegraphics[scale=0.20]{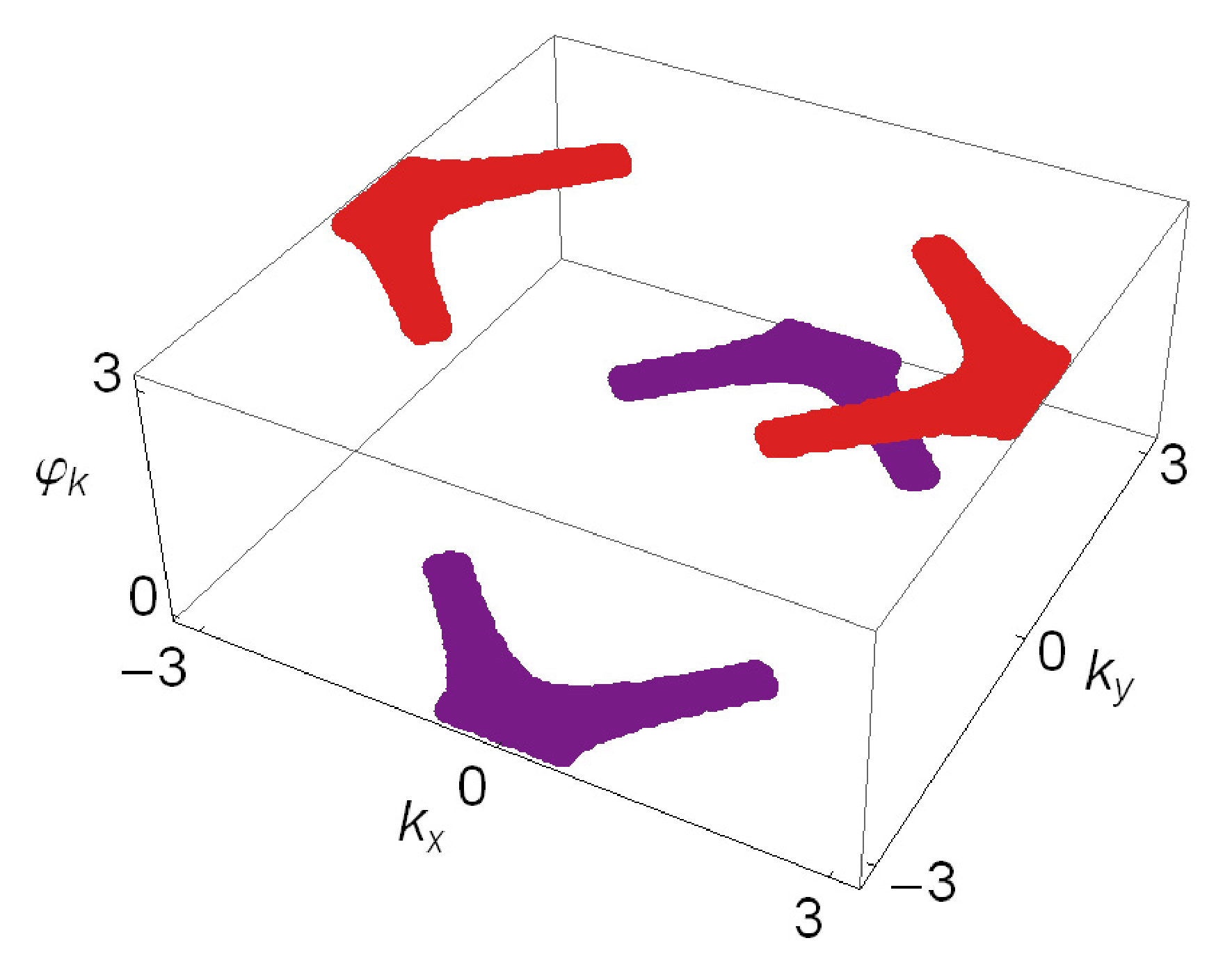}
\caption{The phase of the anomalous thermal average close to the Fermi energy. We have assumed: $V=0.02t$ and $\omega_{0}=0.3t$.}
\label{f2}
\end{figure}
\begin{figure*}[t]
\includegraphics[scale=0.20]{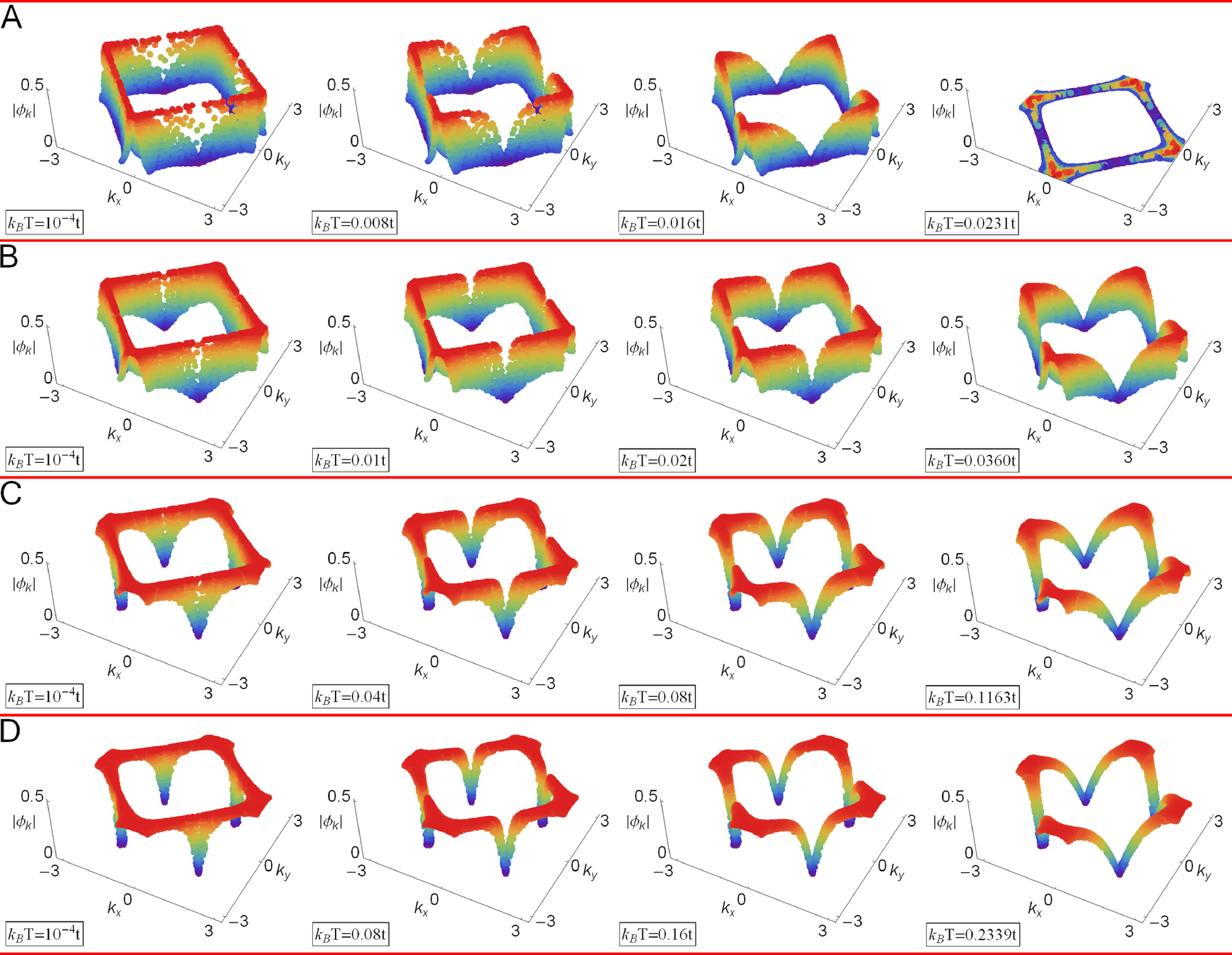}
\caption{ The amplitude of the anomalous thermal average close to the Fermi energy for selected values of $U$ and the temperature. In particular, the figure (A) is for $U=0$, (B) for $U=0.02$t, (C) for $U=0.05$t, and (D) for $U=0.1$t. We have assumed: $V=0.02t$ and $\omega_{0}=0.3t$.}
\label{f3}
\end{figure*}

The Green function for the superconducting state has been calculated using the operator \eq{r1} \cite{Gasser1999A}. The result takes the form:
\begin{equation}
\label{r7}
\left<\left<c_{\kvec\uparrow}|c_{-\kvec\downarrow}\right>\right>_{i\omega_{n}}=
\frac{\frac{1}{N_{0}}\sum^{\omega_{0}}_{\kvec_{2}}W_{\kvec\kvec_{2}}\phi_{\kvec_{2}}}{\omega_{n}^{2}+\varepsilon^{2}_{\kvec}+M^{2}_{\kvec}},
\end{equation}
where $W_{\kvec_{1}\kvec_{2}}$ can be written as: 
$W_{\kvec_{1}\kvec_{2}}\equiv V_{\kvec_{1}\kvec_{2}}+\frac{1}{N^{2}_{0}}\sum^{\omega_{0}}_{\kvec_{3}\kvec_{4}}U_{\kvec_{1}\sim\kvec_{4}}
\phi_{\kvec_{3}}\phi^{\star}_{\kvec_{4}}$, and $\phi_{\kvec}\equiv\phi_{\kvec\downarrow}$. The symbol: 
$\omega_{n}\equiv \frac{\pi}{\beta}\left(2n-1\right)$ denotes the $n$-th fermionic Matsubara frequency; $\beta\equiv 1/k_{B}T$, where $k_{B}$ is the Boltzmann constant. The quantity $M^{2}_{\kvec}$ takes the form:
%
%\begin{widetext}
%
\begin{eqnarray}
\label{r8}
M^{2}_{\kvec}&\equiv&\left(\frac{1}{N_{0}}\sum^{\omega_{0}}_{\kvec_{1}}W_{\kvec_{1}\kvec}\phi^{\star}_{\kvec_{1}}\right)
\left(\frac{1}{N_{0}}\sum^{\omega_{0}}_{\kvec_{2}}W_{\kvec\kvec_{2}}\phi_{\kvec_{2}}\right)\\ \nonumber
&=&
\eta^{2}\left({\bf k}\right)\left(\frac{1}{N_{0}}\sum^{\omega_{0}}_{\kvec_{1}}\eta_{\kvec_{1}}\phi^{\star}_{\kvec_{1}}\right)
\left(\frac{1}{N_{0}}\sum^{\omega_{0}}_{\kvec_{2}}\eta_{\kvec_{2}}\phi_{\kvec_{2}}\right)\\ \nonumber
&\times&
\left[V+U\left(\frac{1}{N_{0}}\sum^{\omega_{0}}_{\kvec_{3}}\eta_{\kvec_{3}}\phi_{\kvec_{3}}\right)
\left(\frac{1}{N_{0}}\sum^{\omega_{0}}_{\kvec_{4}}\eta_{\kvec_{4}}\phi^{\star}_{\kvec_{4}}\right)\right]^{2}.
\end{eqnarray}
%
%\end{widetext}
%

On the basis of Eq.\eq{r7}, we have derived the thermodynamic equation for the anomalous thermal average:
\begin{eqnarray}
\label{r9}
\phi_{\kvec}&=&
\left(\frac{1}{N_{0}}\sum^{\omega_{0}}_{\kvec_{1}}\eta\left(\bf k_{1}\right)\phi_{\kvec_{1}}\right)\\ \nonumber
&\times&
\left[V+U\left(\frac{1}{N_{0}}\sum^{\omega_{0}}_{\kvec_{2}}\eta\left(\bf k_{2}\right)\phi_{\kvec_{2}}\right)
\left(\frac{1}{N_{0}}\sum^{\omega_{0}}_{\kvec_{3}}\eta\left(\bf k_{3}\right)\phi^{\star}_{\kvec_{3}}\right)\right]\\ \nonumber
&\times&
\eta\left(\bf {k}\right)\chi_{\kvec},
\end{eqnarray}
where:
\begin{equation}
\label{r10}
\chi_{\kvec}\equiv\frac{\tan\left[\frac{i\beta}{2}\sqrt{\varepsilon^{2}_{\kvec}+M^{2}_{\kvec}}\right]}
{2i\sqrt{\varepsilon^{2}_{\kvec}+M^{2}_{\kvec}}}.
\end{equation}
We have noticed that Eq.\eq{r9} has been obtained with help of the relation: 
$\frac{1}{\beta}\sum_{n} \left<\left<B|A\right>\right>_{i\omega_{n}}=\left<AB\right>$. In particular: $\frac{1}{\beta}\sum_{n}\frac{1}{\omega^{2}_{n}+\varepsilon^{2}_{\kvec}+M^{2}_{\kvec}}=\frac{1}{2i\sqrt{\varepsilon^{2}_{\kvec}+M^{2}_{\kvec}}}
\tan\left(\frac{i\beta}{2}\sqrt{\varepsilon^{2}_{\kvec}+M^{2}_{\kvec}}\right)$.

The Eq.\eq{r9} has been solved for $7000$ points close to the Fermi energy. The numerical result can be obtained with help of the self-consistent iterative (SCI) procedure originally used for solving the Eliashberg equations defined on the imaginary axis \cite{Szczesniak2012K}, \cite{Szczesniak2012M}. We want to clearly underline that the SCI procedure is appropriate for the large sized system of the non-linear equations. The stability of the numerical solutions is ensured by replacing the initial values with the calculated values and repeatedly running the computer program. The computational errors can be controlled by solving the set independently of each other in the framework of the Fortran and C programming language. 

In Fig.\fig{f1}, we have presented the amplitude of the anomalous thermal average ($|\phi_{\kvec}|$) close to the Fermi energy for different values of the potential $U$. The temperature takes the very low value ($k_{B}T=10^{-4}t$). The obtained results show that beyond the nodal regions the low values of $|\phi_{\kvec}|$ strongly increase with the growth of $U$. For high values of $U$, the observed transition from the $d$-wave BCS to non-BCS behavior results in achieving the nearly-constant value of the amplitude in wide surroundings of the antinodal regions. 
In the nodal regions $|\phi_{\kvec}|$ a wide range of values is assumed.
     
In Fig.\fig{f2}, we have plotted the phase of the anomalous thermal average ($\varphi_{\kvec}$). In contrast to the amplitude $|\phi_{\kvec}|$, the phase is independent from the parameter $U$. Additionally, it has been shown that $\varphi_{\kvec}$ assumes two different values. 

The temperature dependence of $|\phi_{\kvec}|$ for selected values of $U$ has been presented in Fig.\fig{f3}. In the first step, the case of $U=0$ has been considered (Fig.\fig{f3} (A)). It is easy to see that the amplitude decreases with the grow of the temperature and disappears at $T_{C}\simeq 0.0231t$. In other cases ( Fig.\fig{f3} (B)-(D), where $U\neq 0$), the temperature dependence of $|\phi_{\kvec}|$ sharply differs from the prediction based on the pure $d$-wave model. In particular, for $0<T<T_{C}$ the antinodal regions of $|\phi_{\kvec}|$ are very weakly temperature dependent and above the critical temperature extend into the anomalous normal state to the pseudogap temperature. In contrast to the behavior of the antinodal values, the nodal regions of the amplitude strongly decrease with the growth of the temperature 
and also disappears at $T_{C}\simeq 0.0231t$.

\begin{figure}
\includegraphics[scale=0.16]{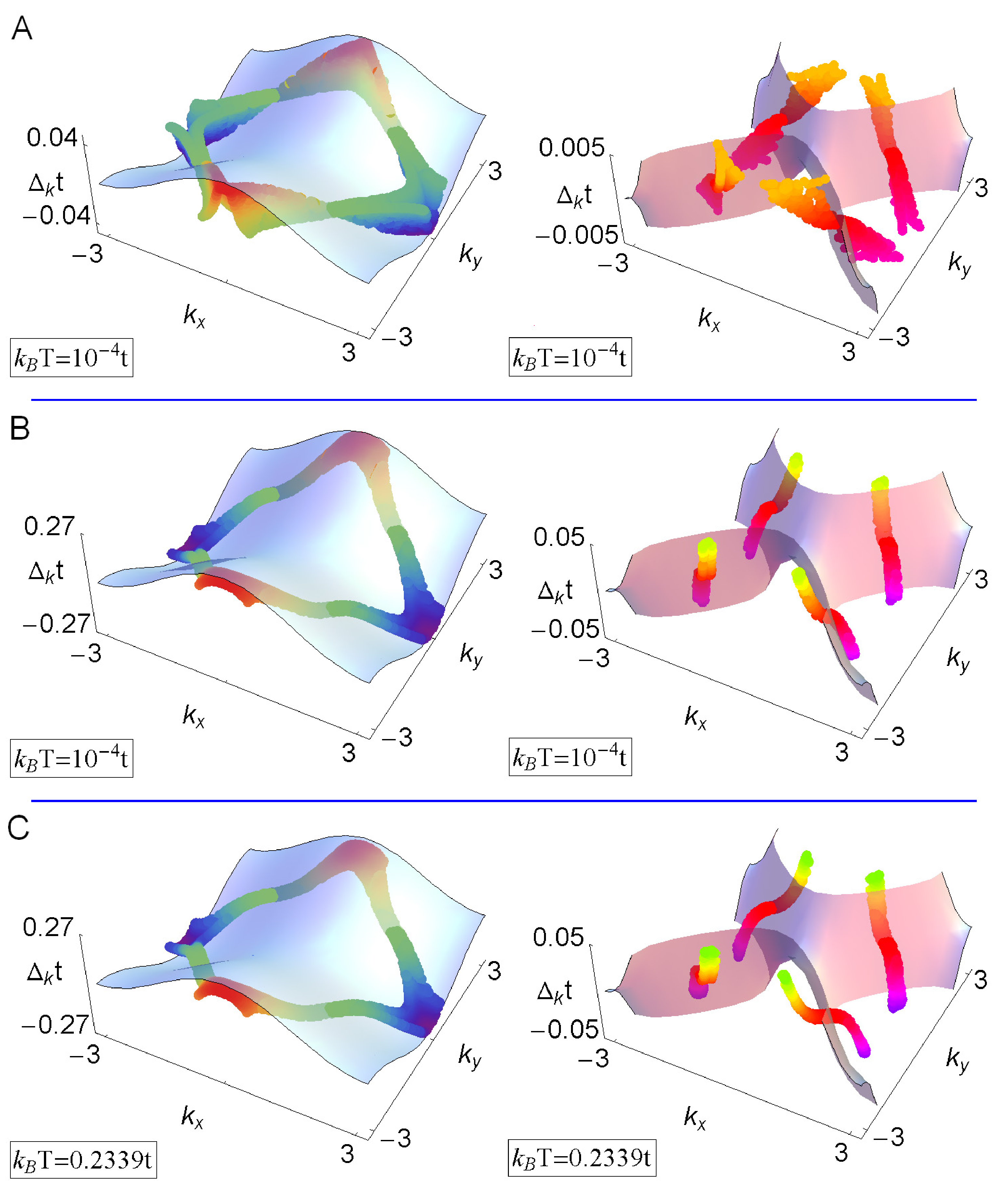}
\caption{The gap parameter close to the Fermi energy for selected values of $U$ and the temperature. In particular, the figure (A) is for $U=0$, (B) and (C) for $U=0.1$t. We have assumed: $V=0.02t$ and $\omega_{0}=0.3t$. The figures on the left show the general form of the gap; the blue surface represents the simple $d$-wave gap. The shape of the gap parameter in the nodal regions has been presented in detail in the figures on the right; the pink surface represents the simple $d$-wave gap.}
\label{f4}
\end{figure}

Now, we can calculate the gap parameter: 
$\Delta_{\kvec}\equiv \eta\left(\bf{k}\right)\left|\phi_{\kvec}\right|\left[V+U\left|\eta\left(\bf{k}\right)\right|\left|\phi_{\kvec}\right|^{2}\right]$.
The form of the function $\Delta_{\kvec}$ has been presented in Fig.\fig{f4}. In particular, we have selected the gap parameter for $U=0$, close to the zero temperature ($k_{B}T=10^{-4}t$), and the gap in the range of high value of the electron-electron-phonon potential ($U=0.1t$) for the two following temperatures: $k_{B}T=10^{-4}t$ and $k_{B}T=0.2339t$. 

\begin{figure*}
\centering
\includegraphics[scale=0.68]{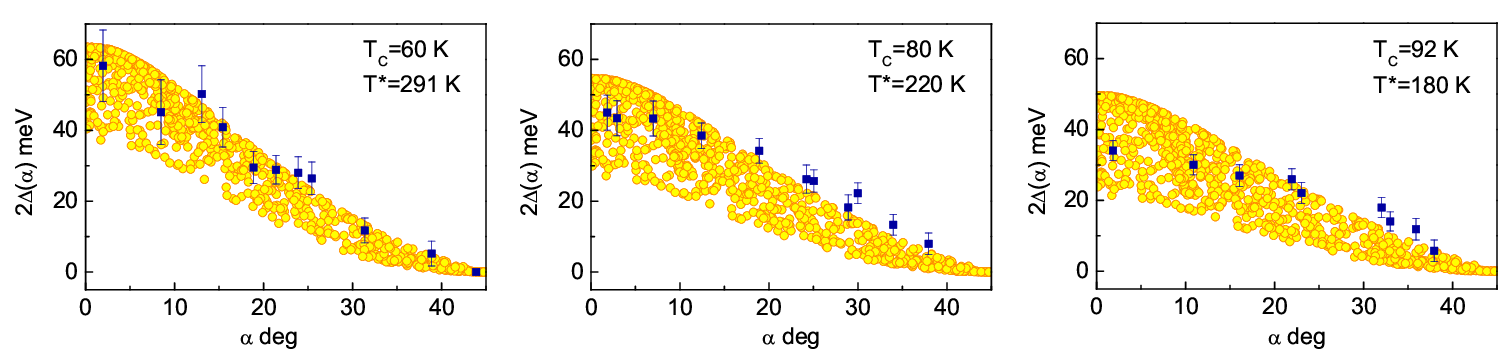}
\caption{The dependence of the energy gap ($2\Delta$) on the angle for the ${\rm YBa_{2}Cu_{3}O_{7-\delta}}$ compound ($T=10$ K). 
The circles represent the theoretical results. The squares are the experimental data \cite{Nakayama2009A}.}
\label{f5}
\end{figure*}

In the case of $U=0$, one can easily see that the gap parameter very well reconstruct the shape of the simple $d$-wave gap 
($\Delta^{\left(\eta\right)}_{\kvec}\equiv\Delta\eta\left(\bf{k}\right)$). 
However, the function $\Delta_{\kvec}$ possesses an enough complicated structure which results from the dependence of the gap amplitude on the wave vector (Fig.\fig{f4} (A)).
We would like to also notice that in the nodal region the gap closely adjoins to the $\Delta^{\left(\eta\right)}_{\kvec}$ surface. Of course, the function $\Delta_{\kvec}$ decreases in the typical way with the increase of the temperature and assumes the zero value at $T_{C}\simeq 0.0231t$.  

For the high value of the electron-electron-phonon potential and $k_{B}T=10^{-4}t$, the amplitude of the gap strong increases and the internal structure of $\Delta_{\kvec}$ becomes much straighter (see Fig.\fig{f4} (B)). In the nodal region the gap agrees relatively well with the $\Delta^{\left(\eta\right)}_{\kvec}$ surface.

In the case of $U=0.1t$ and $k_{B}T=0.2339t$, one can observe anomalous behavior of the gap (Fig. \fig{f4} (C)). In the antinodal region $\Delta_{\kvec}$ has comparable values as for $k_{B}T=10^{-4}t$. From the physical point of view the obtained result indicates that the pseudogap exists in the antinodal region. On the other hand, in the nodal region the gap parameter forms the characteristic area of the very low values. 

The model has been tested on the example of the ${\rm YBa_{2}Cu_{3}O_{7-\delta}}$ compound \cite{Nakayama2009A}. It has been assumed that the nearest-neighbor hopping integral and the characteristic phonon frequency take the values $250$ meV and $75$ meV, respectively \cite{Nunner1999A}, \cite{Bohnen2003A}. 

On the basis of the relations: $T_{C}=T_{C}\left(V\right)$ and $T^{\star}=T^{\star}\left(V,U\right)$, where $T_{C}$ denotes the critical temperature and $T^{\star}$ is the pseudogap temperature, the pairing potentials have been uniquely determined. Then, the equation \eq{r9} has been solved, and the values of $2\Delta\left(\alpha\right)$ have been calculated, wherein $\alpha\equiv\arctan\left({k_{y}}/{k_{x}}\right)$. 

We notice that in the presented approach, the finite doping is included in the values of the pairing potentials V and U, because $T_{C}$ and $T^{\star}$ are the functions of doping. The presented method of analysis allows the partial simulation of the influence of the chemical potential on the energy gap. 

We boldly underline that the rigorous theory should possess the equation for the chemical potential, since the correlated band is narrow and practically all electrons participate in pairing.  However, in this case the extremely complex numerical calculations should be made (in the Eliashberg approach the summation by ${\bf k}$ and the Matsubara frequencies must be taken into consideration).

The theoretical and the experimental results have been compared with each other in Fig. \fig{f5}. It can be easily seen that the presented model in the correct way predicts the course of the experimental data.  

\vspace*{0.25cm}

In the paper, we have tested the microscopic model for the high-$T_{C}$ superconductivity, which is based on the electron-phonon and electron-electron-phonon interaction. In particular, the structure of the gap parameter has been studied. The presented results have been obtained using the thermodynamic equation for the anomalous thermal average. 

It has been stated that for $U=0$, the gap parameter very well reconstruct the $d$-wave behavior. However, $\Delta_{\kvec}$ possesses a complicated internal structure. 

In the opposite case (the large value of $U$) the behavior of the gap parameter is anomalous. In particular, the antinodal part of $\Delta_{\kvec}$ is very weakly temperature dependent and above $T_{C}$ it extends into the anomalous normal state to the pseudogap temperature. On the other hand, the nodal part is temperature dependent and for $T>T_{C}$ the values of the gap disappear. The above results suggests that the pseudogap in the hole-doped cuprates is connected with the antinodal part of the energy gap. 

In the last paragraph, let us notice that the presented model has been tested for the ${\rm YBa_{2}Cu_{3}O_{7-\delta}}$ compound. A good agreement between the theoretical predictions and the experimental results has been achieved.

In the future, we will study the relative stability of the $s$- or the extended-$s$-wave form of the fermionic order parameter. This issue is especially interesting in the case of the electron-doped cuprates, where the symmetry of $\Delta_{\kvec}$ is unclear.

%
%%%%%%%%%%%%%%%%%%%%%%%%%%%%%%%%%%%%%%%%%%%%%%%%%%%%%%%%%%%%%%%%%%%%%%%%%%%%%%%
\begin{acknowledgments}

The authors would like to thank Prof. K. Dzili{\'n}ski for providing excellent working conditions and financial support.\\ 

Computational calculations were performed at the Poznan Supercomputing and Networking Center.
\end{acknowledgments}
%%%%%%%%%%%%%%%%%%%%%%%%%%%%%%%%%%%%%%%%%%%%%%%%%%%%%%%%%%%%%%%%%%%%%%%%%%%%%%%
%
%(Literatura)%%%%%%%%%%

%%%%%%%%%%%%%%%%%%%%%%%
%

\begin{thebibliography}{45}
\expandafter\ifx\csname natexlab\endcsname\relax\def\natexlab#1{#1}\fi
\expandafter\ifx\csname bibnamefont\endcsname\relax
  \def\bibnamefont#1{#1}\fi
\expandafter\ifx\csname bibfnamefont\endcsname\relax
  \def\bibfnamefont#1{#1}\fi
\expandafter\ifx\csname citenamefont\endcsname\relax
  \def\citenamefont#1{#1}\fi
\expandafter\ifx\csname url\endcsname\relax
  \def\url#1{\texttt{#1}}\fi
\expandafter\ifx\csname urlprefix\endcsname\relax\def\urlprefix{URL }\fi
\providecommand{\bibinfo}[2]{#2}
\providecommand{\eprint}[2][]{\url{#2}}

\bibitem[{\citenamefont{Bednorz and M{\"u}ller}(1986)}]{Bednorz1986A}
\bibinfo{author}{\bibfnamefont{J.~G.} \bibnamefont{Bednorz}} \bibnamefont{and}
  \bibinfo{author}{\bibfnamefont{K.~A.} \bibnamefont{M{\"u}ller}},
  \bibinfo{journal}{Z. Phys. B} \textbf{\bibinfo{volume}{64}},
  \bibinfo{pages}{189} (\bibinfo{year}{1986}).

\bibitem[{\citenamefont{Bednorz and M{\"u}ller}(1988)}]{Bednorz1988A}
\bibinfo{author}{\bibfnamefont{J.~G.} \bibnamefont{Bednorz}} \bibnamefont{and}
  \bibinfo{author}{\bibfnamefont{K.~A.} \bibnamefont{M{\"u}ller}},
  \bibinfo{journal}{Rev. Mod. Phys.} \textbf{\bibinfo{volume}{60}},
  \bibinfo{pages}{585} (\bibinfo{year}{1988}).

\bibitem[{\citenamefont{Dagotto}(1994)}]{Dagotto1994A}
\bibinfo{author}{\bibfnamefont{E.}~\bibnamefont{Dagotto}},
  \bibinfo{journal}{Rev. Mod. Phys.} \textbf{\bibinfo{volume}{66}},
  \bibinfo{pages}{763} (\bibinfo{year}{1994}).

\bibitem[{\citenamefont{Harlingen}(1995)}]{Harlingen1995A}
\bibinfo{author}{\bibfnamefont{D.~J.~V.} \bibnamefont{Harlingen}},
  \bibinfo{journal}{Rev. Mod. Phys.} \textbf{\bibinfo{volume}{67}},
  \bibinfo{pages}{515535} (\bibinfo{year}{1995}).

\bibitem[{\citenamefont{Tsuei and Kirtley}(2000{\natexlab{a}})}]{Tsuei2000A}
\bibinfo{author}{\bibfnamefont{C.~C.} \bibnamefont{Tsuei}} \bibnamefont{and}
  \bibinfo{author}{\bibfnamefont{J.~R.} \bibnamefont{Kirtley}},
  \bibinfo{journal}{Rev. Mod. Phys.} \textbf{\bibinfo{volume}{72}},
  \bibinfo{pages}{969} (\bibinfo{year}{2000}{\natexlab{a}}).

\bibitem[{\citenamefont{Smilde et~al.}(2005)\citenamefont{Smilde, Golubov,
  Ariando, Rijnders, Dekkers, Harkema, Blank, Rogalla, and
  Hilgenkamp}}]{Smilde2005A}
\bibinfo{author}{\bibfnamefont{H.~J.~H.} \bibnamefont{Smilde}},
  \bibinfo{author}{\bibfnamefont{A.~A.} \bibnamefont{Golubov}},
  \bibinfo{author}{\bibnamefont{Ariando}},
  \bibinfo{author}{\bibfnamefont{G.}~\bibnamefont{Rijnders}},
  \bibinfo{author}{\bibfnamefont{J.~M.} \bibnamefont{Dekkers}},
  \bibinfo{author}{\bibfnamefont{S.}~\bibnamefont{Harkema}},
  \bibinfo{author}{\bibfnamefont{D.~H.~A.} \bibnamefont{Blank}},
  \bibinfo{author}{\bibfnamefont{H.}~\bibnamefont{Rogalla}}, \bibnamefont{and}
  \bibinfo{author}{\bibfnamefont{H.}~\bibnamefont{Hilgenkamp}},
  \bibinfo{journal}{Phys. Rev. Lett.} \textbf{\bibinfo{volume}{95}},
  \bibinfo{pages}{257001} (\bibinfo{year}{2005}).

\bibitem[{\citenamefont{Armitage et~al.}(2010)\citenamefont{Armitage, Fournier,
  and Greene}}]{Armitage2010A}
\bibinfo{author}{\bibfnamefont{N.~P.} \bibnamefont{Armitage}},
  \bibinfo{author}{\bibfnamefont{P.}~\bibnamefont{Fournier}}, \bibnamefont{and}
  \bibinfo{author}{\bibfnamefont{R.~L.} \bibnamefont{Greene}},
  \bibinfo{journal}{Rev. Mod. Phys.} \textbf{\bibinfo{volume}{82}},
  \bibinfo{pages}{2421} (\bibinfo{year}{2010}).

\bibitem[{\citenamefont{Kokales et~al.}(2000)\citenamefont{Kokales, Fournier,
  Mercaldo, Talanov, Greene, and Anlage}}]{Kokales2000A}
\bibinfo{author}{\bibfnamefont{J.~D.} \bibnamefont{Kokales}},
  \bibinfo{author}{\bibfnamefont{P.}~\bibnamefont{Fournier}},
  \bibinfo{author}{\bibfnamefont{L.~V.} \bibnamefont{Mercaldo}},
  \bibinfo{author}{\bibfnamefont{V.~V.} \bibnamefont{Talanov}},
  \bibinfo{author}{\bibfnamefont{R.~L.} \bibnamefont{Greene}},
  \bibnamefont{and} \bibinfo{author}{\bibfnamefont{S.~M.}
  \bibnamefont{Anlage}}, \bibinfo{journal}{Phys. Rev. Lett.}
  \textbf{\bibinfo{volume}{85}}, \bibinfo{pages}{3696} (\bibinfo{year}{2000}).

\bibitem[{\citenamefont{Prozorov et~al.}(2000)\citenamefont{Prozorov,
  Giannetta, Fournier, and Greene}}]{Prozorov2000A}
\bibinfo{author}{\bibfnamefont{R.}~\bibnamefont{Prozorov}},
  \bibinfo{author}{\bibfnamefont{R.~W.} \bibnamefont{Giannetta}},
  \bibinfo{author}{\bibfnamefont{P.}~\bibnamefont{Fournier}}, \bibnamefont{and}
  \bibinfo{author}{\bibfnamefont{R.~L.} \bibnamefont{Greene}},
  \bibinfo{journal}{Phys. Rev. Lett.} \textbf{\bibinfo{volume}{85}},
  \bibinfo{pages}{3700} (\bibinfo{year}{2000}).

\bibitem[{\citenamefont{Tsuei and Kirtley}(2000{\natexlab{b}})}]{Tsuei2000B}
\bibinfo{author}{\bibfnamefont{C.~C.} \bibnamefont{Tsuei}} \bibnamefont{and}
  \bibinfo{author}{\bibfnamefont{J.~R.} \bibnamefont{Kirtley}},
  \bibinfo{journal}{Phys. Rev. Lett.} \textbf{\bibinfo{volume}{85}},
  \bibinfo{pages}{182185} (\bibinfo{year}{2000}{\natexlab{b}}).

\bibitem[{\citenamefont{Armitage et~al.}(2001)\citenamefont{Armitage, Lu, Feng,
  Kim, Damascelli, Shen, Ronning, Shen, Onose, Taguchi et~al.}}]{Armitage2001A}
\bibinfo{author}{\bibfnamefont{N.~P.} \bibnamefont{Armitage}},
  \bibinfo{author}{\bibfnamefont{D.~H.} \bibnamefont{Lu}},
  \bibinfo{author}{\bibfnamefont{D.~L.} \bibnamefont{Feng}},
  \bibinfo{author}{\bibfnamefont{C.}~\bibnamefont{Kim}},
  \bibinfo{author}{\bibfnamefont{A.}~\bibnamefont{Damascelli}},
  \bibinfo{author}{\bibfnamefont{K.~M.} \bibnamefont{Shen}},
  \bibinfo{author}{\bibfnamefont{F.}~\bibnamefont{Ronning}},
  \bibinfo{author}{\bibfnamefont{Z.-X.} \bibnamefont{Shen}},
  \bibinfo{author}{\bibfnamefont{Y.}~\bibnamefont{Onose}},
  \bibinfo{author}{\bibfnamefont{Y.}~\bibnamefont{Taguchi}},
  \bibnamefont{et~al.}, \bibinfo{journal}{Phys. Rev. Lett.}
  \textbf{\bibinfo{volume}{86}}, \bibinfo{pages}{1126} (\bibinfo{year}{2001}).

\bibitem[{\citenamefont{Sato et~al.}(2001)\citenamefont{Sato, Kamiyama,
  Takahashi, Kurahashi, and Yamada}}]{Sato2001A}
\bibinfo{author}{\bibfnamefont{T.}~\bibnamefont{Sato}},
  \bibinfo{author}{\bibfnamefont{T.}~\bibnamefont{Kamiyama}},
  \bibinfo{author}{\bibfnamefont{T.}~\bibnamefont{Takahashi}},
  \bibinfo{author}{\bibfnamefont{K.}~\bibnamefont{Kurahashi}},
  \bibnamefont{and} \bibinfo{author}{\bibfnamefont{K.}~\bibnamefont{Yamada}},
  \bibinfo{journal}{Science} \textbf{\bibinfo{volume}{291}},
  \bibinfo{pages}{1517} (\bibinfo{year}{2001}).

\bibitem[{\citenamefont{Blumberg et~al.}(2002)\citenamefont{Blumberg, Koitzsch,
  Gozar, Dennis, Kendziora, Fournier, and Greene}}]{Blumberg2002A}
\bibinfo{author}{\bibfnamefont{G.}~\bibnamefont{Blumberg}},
  \bibinfo{author}{\bibfnamefont{A.}~\bibnamefont{Koitzsch}},
  \bibinfo{author}{\bibfnamefont{A.}~\bibnamefont{Gozar}},
  \bibinfo{author}{\bibfnamefont{B.~S.} \bibnamefont{Dennis}},
  \bibinfo{author}{\bibfnamefont{C.~A.} \bibnamefont{Kendziora}},
  \bibinfo{author}{\bibfnamefont{P.}~\bibnamefont{Fournier}}, \bibnamefont{and}
  \bibinfo{author}{\bibfnamefont{R.~L.} \bibnamefont{Greene}},
  \bibinfo{journal}{Phys. Rev. Lett.} \textbf{\bibinfo{volume}{88}},
  \bibinfo{pages}{107002} (\bibinfo{year}{2002}).

\bibitem[{\citenamefont{Biswas et~al.}(2002)\citenamefont{Biswas, Fournier,
  Qazilbash, Smolyaninova, Balci, and Greene}}]{Biswas2002A}
\bibinfo{author}{\bibfnamefont{A.}~\bibnamefont{Biswas}},
  \bibinfo{author}{\bibfnamefont{P.}~\bibnamefont{Fournier}},
  \bibinfo{author}{\bibfnamefont{M.~M.} \bibnamefont{Qazilbash}},
  \bibinfo{author}{\bibfnamefont{V.~N.} \bibnamefont{Smolyaninova}},
  \bibinfo{author}{\bibfnamefont{H.}~\bibnamefont{Balci}}, \bibnamefont{and}
  \bibinfo{author}{\bibfnamefont{R.~L.} \bibnamefont{Greene}},
  \bibinfo{journal}{Phys. Rev. Lett.} \textbf{\bibinfo{volume}{88}},
  \bibinfo{pages}{207004} (\bibinfo{year}{2002}).

\bibitem[{\citenamefont{Skinta et~al.}(2002)\citenamefont{Skinta, Kim,
  Lemberger, Greibe, and Naito}}]{Skinta2002A}
\bibinfo{author}{\bibfnamefont{J.~A.} \bibnamefont{Skinta}},
  \bibinfo{author}{\bibfnamefont{M.~S.} \bibnamefont{Kim}},
  \bibinfo{author}{\bibfnamefont{T.~R.} \bibnamefont{Lemberger}},
  \bibinfo{author}{\bibfnamefont{T.}~\bibnamefont{Greibe}}, \bibnamefont{and}
  \bibinfo{author}{\bibfnamefont{M.}~\bibnamefont{Naito}},
  \bibinfo{journal}{Phys. Rev. Lett.} \textbf{\bibinfo{volume}{88}},
  \bibinfo{pages}{207005} (\bibinfo{year}{2002}).

\bibitem[{\citenamefont{Damascelli et~al.}(2003)\citenamefont{Damascelli,
  Hussain, and Shen}}]{Damascelli2003A}
\bibinfo{author}{\bibfnamefont{A.}~\bibnamefont{Damascelli}},
  \bibinfo{author}{\bibfnamefont{Z.}~\bibnamefont{Hussain}}, \bibnamefont{and}
  \bibinfo{author}{\bibfnamefont{Z.~-X.} \bibnamefont{Shen}},
  \bibinfo{journal}{Rev. Mod. Phys.} \textbf{\bibinfo{volume}{75}},
  \bibinfo{pages}{473} (\bibinfo{year}{2003}).

%%%%%%%%%%%%%%%%%%%%%%%%%%%%%%%%%%%%%%%%%%%%%%%%%%%%%%%%%%%%%%%%%%%%%%%%%%%%%%%%%%%%%%%%%%%%%%%%%%%%%%%%%%%%%%%%%%%%%%%%%%%%%%%%%%%%%%%%%%%%%%%%%%%%%%%%%%%%%

\bibitem[{\citenamefont{Yoshida et~al.}(2012)\citenamefont{Yoshida, Hashimoto, Vishik, Shen, and Fujimori}}]{Yoshida2012A}
  \bibinfo{author}{\bibfnamefont{T.}~\bibnamefont{Yoshida}},
  \bibinfo{author}{\bibfnamefont{M.}~\bibnamefont{Hashimoto}}, 
  \bibinfo{author}{\bibfnamefont{I.~M.}~\bibnamefont{Vishik}},
  \bibinfo{author}{\bibfnamefont{Z.~-X.}~\bibnamefont{Shen}}, 
  \bibnamefont{and}
  \bibinfo{author}{\bibfnamefont{A.}~\bibnamefont{Fujimori}},
  \bibinfo{journal}{J. Phys. Soc. Jpn.} \textbf{\bibinfo{volume}{81}},
  \bibinfo{pages}{011006} (\bibinfo{year}{2012}).

\bibitem[{\citenamefont{Hashimoto et~al.}(2012)\citenamefont{Hashimoto, He, Vishik, Schmitt, Moore, Lu, Yoshida, Eisaki, Hussain, Devereaux, and Shen}}]{Hashimoto2012A}
  \bibinfo{author}{\bibfnamefont{M.}~\bibnamefont{Hashimoto}},
  \bibinfo{author}{\bibfnamefont{R.~-H.}~\bibnamefont{He}}, 
  \bibinfo{author}{\bibfnamefont{I.~M.}~\bibnamefont{Vishik}},
  \bibinfo{author}{\bibfnamefont{F.}~\bibnamefont{Schmitt}},
  \bibinfo{author}{\bibfnamefont{R.~G.}~\bibnamefont{Moore}},
  \bibinfo{author}{\bibfnamefont{D.~H.}~\bibnamefont{Lu}}, 
  \bibinfo{author}{\bibfnamefont{Y.}~\bibnamefont{Yoshida}},
  \bibinfo{author}{\bibfnamefont{H.}~\bibnamefont{Eisaki}},
  \bibinfo{author}{\bibfnamefont{Z.}~\bibnamefont{Hussain}},
  \bibinfo{author}{\bibfnamefont{T.~P.}~\bibnamefont{Devereaux}}, 
  \bibnamefont{and}
  \bibinfo{author}{\bibfnamefont{Z.~-X.}~\bibnamefont{Shen}},
  \bibinfo{journal}{Phys. Rev. B} \textbf{\bibinfo{volume}{86}},
  \bibinfo{pages}{094504} (\bibinfo{year}{2012}).

\bibitem[{\citenamefont{Hashimoto et~al.}(2014)\citenamefont{Hashimoto, Vishik, He, Devereaux, and Shen}}]{Hashimoto2014A}
  \bibinfo{author}{\bibfnamefont{M.}~\bibnamefont{Hashimoto}}, 
  \bibinfo{author}{\bibfnamefont{I.~M.}~\bibnamefont{Vishik}},
  \bibinfo{author}{\bibfnamefont{R.~-H.}~\bibnamefont{He}},
  \bibinfo{author}{\bibfnamefont{T.~P.}~\bibnamefont{Devereaux}},
  \bibnamefont{and}
  \bibinfo{author}{\bibfnamefont{Z.~-X.}~\bibnamefont{Shen}}, 
  \bibinfo{journal}{Nature Physics} \textbf{\bibinfo{volume}{10}},
  \bibinfo{pages}{483} (\bibinfo{year}{2014}).

\bibitem[{\citenamefont{Vishik et~al.}(2014)\citenamefont{Vishik, Barisic, Chan, Li, Xia, Yu, Zhao, Lee, Meevasana, Devereaux, Greven, and Shen,}}]{Vishik2014A}
  \bibinfo{author}{\bibfnamefont{I.~M.}~\bibnamefont{Vishik}},
  \bibinfo{author}{\bibfnamefont{N.}~\bibnamefont{Barisic}},
  \bibinfo{author}{\bibfnamefont{M.~K.}~\bibnamefont{Chan}},
  \bibinfo{author}{\bibfnamefont{Y.}~\bibnamefont{Li}},
  \bibinfo{author}{\bibfnamefont{D.~D.}~\bibnamefont{Xia}},
  \bibinfo{author}{\bibfnamefont{G.}~\bibnamefont{Yu}},
  \bibinfo{author}{\bibfnamefont{X.}~\bibnamefont{Zhao}},
  \bibinfo{author}{\bibfnamefont{W.~S.}~\bibnamefont{Lee}},
  \bibinfo{author}{\bibfnamefont{W.}~\bibnamefont{Meevasana}},
  \bibinfo{author}{\bibfnamefont{T.~P.}~\bibnamefont{Devereaux}},
  \bibinfo{author}{\bibfnamefont{M.}~\bibnamefont{Greven}},
  \bibnamefont{and}
  \bibinfo{author}{\bibfnamefont{Z.~-X.}~\bibnamefont{Shen}}, 
  \bibinfo{journal}{Phys. Rev. B} \textbf{\bibinfo{volume}{89}},
  \bibinfo{pages}{195141} (\bibinfo{year}{2014}).

%%%%%%%%%%%%%%%%%%%%%%%%%%%%%%%%%%%%%%%%%%%%%%%%%%%%%%%%%%%%%%%%%%%%%%%%%%%%%%%%%%%%%%%%%%%%%%%%%%%%%%%%%%%%%%%%%%%%%%%%%%%%%%%%%%%%%%%%%%%%%%%%%%%%%%%%%%%%%

\bibitem[{\citenamefont{Tanaka et~al.}(2006)\citenamefont{Tanaka, Lee, Lu,
  Fujimori, Fujii, Risdiana, Terasaki, Scalapino, Devereaux, Hussain
  et~al.}}]{Tanaka2006A}
\bibinfo{author}{\bibfnamefont{K.}~\bibnamefont{Tanaka}},
  \bibinfo{author}{\bibfnamefont{W.~S.} \bibnamefont{Lee}},
  \bibinfo{author}{\bibfnamefont{D.~H.} \bibnamefont{Lu}},
  \bibinfo{author}{\bibfnamefont{A.}~\bibnamefont{Fujimori}},
  \bibinfo{author}{\bibfnamefont{T.}~\bibnamefont{Fujii}},
  \bibinfo{author}{\bibnamefont{Risdiana}},
  \bibinfo{author}{\bibfnamefont{I.}~\bibnamefont{Terasaki}},
  \bibinfo{author}{\bibfnamefont{D.~J.} \bibnamefont{Scalapino}},
  \bibinfo{author}{\bibfnamefont{T.~P.} \bibnamefont{Devereaux}},
  \bibinfo{author}{\bibfnamefont{Z.}~\bibnamefont{Hussain}},
  \bibnamefont{et~al.}, \bibinfo{journal}{Science}
  \textbf{\bibinfo{volume}{314}}, \bibinfo{pages}{1910} (\bibinfo{year}{2006}).

\bibitem[{\citenamefont{Lee et~al.}(2007)\citenamefont{Lee, Vishik, Tanaka, Lu,
  Sasagawa, Nagaosa, Devereaux, Hussain, and Shen}}]{Lee2007A}
\bibinfo{author}{\bibfnamefont{W.~S.} \bibnamefont{Lee}},
  \bibinfo{author}{\bibfnamefont{I.~M.} \bibnamefont{Vishik}},
  \bibinfo{author}{\bibfnamefont{K.}~\bibnamefont{Tanaka}},
  \bibinfo{author}{\bibfnamefont{D.~H.} \bibnamefont{Lu}},
  \bibinfo{author}{\bibfnamefont{T.}~\bibnamefont{Sasagawa}},
  \bibinfo{author}{\bibfnamefont{N.}~\bibnamefont{Nagaosa}},
  \bibinfo{author}{\bibfnamefont{T.~P.} \bibnamefont{Devereaux}},
  \bibinfo{author}{\bibfnamefont{Z.}~\bibnamefont{Hussain}}, \bibnamefont{and}
  \bibinfo{author}{\bibfnamefont{Z.~X.} \bibnamefont{Shen}},
  \bibinfo{journal}{Nature (London)} \textbf{\bibinfo{volume}{450}},
  \bibinfo{pages}{81} (\bibinfo{year}{2007}).

\bibitem[{\citenamefont{Kondo et~al.}(2007)\citenamefont{Kondo, Takeuchi,
  Kaminski, Tsuda, and Shin}}]{Kondo2007A}
\bibinfo{author}{\bibfnamefont{T.}~\bibnamefont{Kondo}},
  \bibinfo{author}{\bibfnamefont{T.}~\bibnamefont{Takeuchi}},
  \bibinfo{author}{\bibfnamefont{A.}~\bibnamefont{Kaminski}},
  \bibinfo{author}{\bibfnamefont{S.}~\bibnamefont{Tsuda}}, \bibnamefont{and}
  \bibinfo{author}{\bibfnamefont{S.}~\bibnamefont{Shin}},
  \bibinfo{journal}{Phys. Rev. Lett.} \textbf{\bibinfo{volume}{98}},
  \bibinfo{pages}{267004} (\bibinfo{year}{2007}).

\bibitem[{\citenamefont{Terashima et~al.}(2007)\citenamefont{Terashima, Matsui,
  Sato, Takahashi, Kofu, and Hirota}}]{Terashima2007A}
\bibinfo{author}{\bibfnamefont{K.}~\bibnamefont{Terashima}},
  \bibinfo{author}{\bibfnamefont{H.}~\bibnamefont{Matsui}},
  \bibinfo{author}{\bibfnamefont{T.}~\bibnamefont{Sato}},
  \bibinfo{author}{\bibfnamefont{T.}~\bibnamefont{Takahashi}},
  \bibinfo{author}{\bibfnamefont{M.}~\bibnamefont{Kofu}}, \bibnamefont{and}
  \bibinfo{author}{\bibfnamefont{K.}~\bibnamefont{Hirota}},
  \bibinfo{journal}{Phys. Rev. Lett.} \textbf{\bibinfo{volume}{99}},
  \bibinfo{pages}{017003} (\bibinfo{year}{2007}).

\bibitem[{\citenamefont{Kanigel et~al.}(2007)\citenamefont{Kanigel, Chatterjee,
  Randeria, Norman, Souma, Shi, Li, Raffy, and Campuzano}}]{Kanigel2007A}
\bibinfo{author}{\bibfnamefont{A.}~\bibnamefont{Kanigel}},
  \bibinfo{author}{\bibfnamefont{U.}~\bibnamefont{Chatterjee}},
  \bibinfo{author}{\bibfnamefont{M.}~\bibnamefont{Randeria}},
  \bibinfo{author}{\bibfnamefont{M.~R.} \bibnamefont{Norman}},
  \bibinfo{author}{\bibfnamefont{S.}~\bibnamefont{Souma}},
  \bibinfo{author}{\bibfnamefont{M.}~\bibnamefont{Shi}},
  \bibinfo{author}{\bibfnamefont{Z.~Z.} \bibnamefont{Li}},
  \bibinfo{author}{\bibfnamefont{H.}~\bibnamefont{Raffy}}, \bibnamefont{and}
  \bibinfo{author}{\bibfnamefont{J.~C.} \bibnamefont{Campuzano}},
  \bibinfo{journal}{Phys. Rev. Lett.} \textbf{\bibinfo{volume}{99}},
  \bibinfo{pages}{157001} (\bibinfo{year}{2007}).

\bibitem[{\citenamefont{Shi et~al.}(2008)\citenamefont{Shi, Chang, Pailhes,
  Norman, Campuzano, Mansson, Claesson, Tjernberg, Bendounan, Patthey
  et~al.}}]{Shi2008A}
\bibinfo{author}{\bibfnamefont{M.}~\bibnamefont{Shi}},
  \bibinfo{author}{\bibfnamefont{J.}~\bibnamefont{Chang}},
  \bibinfo{author}{\bibfnamefont{S.}~\bibnamefont{Pailhes}},
  \bibinfo{author}{\bibfnamefont{M.~R.} \bibnamefont{Norman}},
  \bibinfo{author}{\bibfnamefont{J.~C.} \bibnamefont{Campuzano}},
  \bibinfo{author}{\bibfnamefont{M.}~\bibnamefont{Mansson}},
  \bibinfo{author}{\bibfnamefont{T.}~\bibnamefont{Claesson}},
  \bibinfo{author}{\bibfnamefont{O.}~\bibnamefont{Tjernberg}},
  \bibinfo{author}{\bibfnamefont{A.}~\bibnamefont{Bendounan}},
  \bibinfo{author}{\bibfnamefont{L.}~\bibnamefont{Patthey}},
  \bibnamefont{et~al.}, \bibinfo{journal}{Phys. Rev. Lett.}
  \textbf{\bibinfo{volume}{101}}, \bibinfo{pages}{047002}
  (\bibinfo{year}{2008}).

\bibitem[{\citenamefont{Mesot et~al.}(1999)\citenamefont{Mesot, Norman, Ding,
  Randeria, Campuzano, Paramekanti, Fretwell, Kaminski, Takeuchi, Yokoya
  et~al.}}]{Mesot1999A}
\bibinfo{author}{\bibfnamefont{J.}~\bibnamefont{Mesot}},
  \bibinfo{author}{\bibfnamefont{M.~R.} \bibnamefont{Norman}},
  \bibinfo{author}{\bibfnamefont{H.}~\bibnamefont{Ding}},
  \bibinfo{author}{\bibfnamefont{M.}~\bibnamefont{Randeria}},
  \bibinfo{author}{\bibfnamefont{J.~C.} \bibnamefont{Campuzano}},
  \bibinfo{author}{\bibfnamefont{A.}~\bibnamefont{Paramekanti}},
  \bibinfo{author}{\bibfnamefont{H.~M.} \bibnamefont{Fretwell}},
  \bibinfo{author}{\bibfnamefont{A.}~\bibnamefont{Kaminski}},
  \bibinfo{author}{\bibfnamefont{T.}~\bibnamefont{Takeuchi}},
  \bibinfo{author}{\bibfnamefont{T.}~\bibnamefont{Yokoya}},
  \bibnamefont{et~al.}, \bibinfo{journal}{Phys. Rev. Lett.}
  \textbf{\bibinfo{volume}{83}}, \bibinfo{pages}{840843}
  (\bibinfo{year}{1999}).

\bibitem[{\citenamefont{Szcz{\c{e}}{\'s}niak}(2012{\natexlab{a}})}]{Szczesniak2012D}
\bibinfo{author}{\bibfnamefont{R.}~\bibnamefont{Szcz{\c{e}}{\'s}niak}},
  \bibinfo{journal}{PloS ONE} \textbf{\bibinfo{volume}{7}}, \bibinfo{pages}{e31873} (\bibinfo{year}{2012}{\natexlab{a}}).

\bibitem[{\citenamefont{Szcz{\c{e}}{\'s}niak and
  Durajski}(2014)}]{Szczesniak2012E}
\bibinfo{author}{\bibfnamefont{R.}~\bibnamefont{Szcz{\c{e}}{\'s}niak}}
  \bibnamefont{and} \bibinfo{author}{\bibfnamefont{A.~P.}
  \bibnamefont{Durajski}}, \bibinfo{journal}{J. Supercond. Nov. Magn.}
  \textbf{\bibinfo{volume}{27}}, \bibinfo{pages}{1363}
  (\bibinfo{year}{2014}).
  

\bibitem[{\citenamefont{Fr{\"o}hlich}(1950)}]{Frohlich1950A}
\bibinfo{author}{\bibfnamefont{H.}~\bibnamefont{Fr{\"o}hlich}},
  \bibinfo{journal}{Phys. Rev.} \textbf{\bibinfo{volume}{79}},
  \bibinfo{pages}{845} (\bibinfo{year}{1950}).

\bibitem[{\citenamefont{Fr{\"o}hlich}(1954)}]{Frohlich1954A}
\bibinfo{author}{\bibfnamefont{H.}~\bibnamefont{Fr{\"o}hlich}},
  \bibinfo{journal}{Proc. R. Soc. A} \textbf{\bibinfo{volume}{223}},
  \bibinfo{pages}{296} (\bibinfo{year}{1954}).

\bibitem[{\citenamefont{Szcz{\c{e}}{\'s}niak
  et~al.}(2001)\citenamefont{Szcz{\c{e}}{\'s}niak, Mierzejewski, Zieli{\'n}ski,
  and Entel}}]{Szczesniak2001A}
\bibinfo{author}{\bibfnamefont{R.}~\bibnamefont{Szcz{\c{e}}{\'s}niak}},
  \bibinfo{author}{\bibfnamefont{M.}~\bibnamefont{Mierzejewski}},
  \bibinfo{author}{\bibfnamefont{J.}~\bibnamefont{Zieli{\'n}ski}},
  \bibnamefont{and} \bibinfo{author}{\bibfnamefont{P.}~\bibnamefont{Entel}},
  \bibinfo{journal}{Solid State Commun.} \textbf{\bibinfo{volume}{117}},
  \bibinfo{pages}{369} (\bibinfo{year}{2001}).

\bibitem[{\citenamefont{Szcz{\c{e}}{\'s}niak}(2006)}]{Szczesniak2006B}
\bibinfo{author}{\bibfnamefont{R.}~\bibnamefont{Szcz{\c{e}}{\'s}niak}},
  \bibinfo{journal}{Solid State Commun.} \textbf{\bibinfo{volume}{138}},
  \bibinfo{pages}{347} (\bibinfo{year}{2006}).

%%%%%%%%%%%%%%%%%%%%%%%%%%%%%%%%%%%%%%%%%%%%%%%%%%%%%%%O silnych korelacjach elektron-elektron

\bibitem[{\citenamefont{Hubbard}(1963)}]{Hubbard1963A}
\bibinfo{author}{\bibfnamefont{J.}~\bibnamefont{Hubbard}},
  \bibinfo{journal}{Proc. R. Soc. London, Ser. A}
  \textbf{\bibinfo{volume}{276}}, \bibinfo{pages}{238} (\bibinfo{year}{1963}).

\bibitem[{\citenamefont{Emery}(1987)\citenamefont{Emery}}]{Emery1987A}
  \bibinfo{author}{\bibfnamefont{V.~J.}~\bibnamefont{Emery}},  
  \bibinfo{journal}{Phys. Rev. Lett.} \textbf{\bibinfo{volume}{58}},
  \bibinfo{pages}{2794} (\bibinfo{year}{1987}).

\bibitem[{\citenamefont{Littlewood et~al.}(1987)\citenamefont{Littlewood}}]{Littlewood1987A}
  \bibinfo{author}{\bibfnamefont{P.~B.}~\bibnamefont{Littlewood}},
  \bibinfo{author}{\bibfnamefont{C.~M.}~\bibnamefont{Varma}},
  \bibnamefont{and}
  \bibinfo{author}{\bibfnamefont{E.}~\bibnamefont{Abrahams}},
  \bibinfo{journal}{Phys. Rev. Lett.} \textbf{\bibinfo{volume}{60}},
  \bibinfo{pages}{379} (\bibinfo{year}{1987}).

\bibitem[{\citenamefont{Anderson}(1987)\citenamefont{Anderson}}]{Anderson1987A}
  \bibinfo{author}{\bibfnamefont{P.~W.}~\bibnamefont{Anderson}},
  \bibinfo{journal}{Science} \textbf{\bibinfo{volume}{235}},
  \bibinfo{pages}{1196} (\bibinfo{year}{1987}).

\bibitem[{\citenamefont{Millis et~al.}(1990)\citenamefont{Millis}}]{Millis1990A}
  \bibinfo{author}{\bibfnamefont{J.~A.}~\bibnamefont{Millis}},
  \bibinfo{author}{\bibfnamefont{H.}~\bibnamefont{Monien}},
  \bibnamefont{and}
  \bibinfo{author}{\bibfnamefont{D.}~\bibnamefont{Pines}},
  \bibinfo{journal}{Phys. Rev. B} \textbf{\bibinfo{volume}{42}},
  \bibinfo{pages}{167} (\bibinfo{year}{1990}).

\bibitem[{\citenamefont{Monthoux et~al.}(1992)\citenamefont{Monthoux}}]{Monthoux1992A}
  \bibinfo{author}{\bibfnamefont{P.}~\bibnamefont{Monthoux}},
  \bibnamefont{and}
  \bibinfo{author}{\bibfnamefont{D.}~\bibnamefont{Pines}},
  \bibinfo{journal}{Phys. Rev. Lett.} \textbf{\bibinfo{volume}{69}},
  \bibinfo{pages}{961} (\bibinfo{year}{1992}).

\bibitem[{\citenamefont{Lee et~al.}(1987)\citenamefont{Lee}}]{Lee2006A}
  \bibinfo{author}{\bibfnamefont{P.~A.}~\bibnamefont{Lee}},
  \bibinfo{author}{\bibfnamefont{N.}~\bibnamefont{Nagaosa}},
  \bibnamefont{and}
  \bibinfo{author}{\bibfnamefont{C.~-G.}~\bibnamefont{Wen}},
  \bibinfo{journal}{Rev. Mod. Phys.} \textbf{\bibinfo{volume}{78}},
  \bibinfo{pages}{17} (\bibinfo{year}{2006}).

\bibitem[{\citenamefont{Chao et~al.}(1977)\citenamefont{Chao}}]{Chao1977A}
  \bibinfo{author}{\bibfnamefont{K.~A.}~\bibnamefont{Chao}},
  \bibinfo{author}{\bibfnamefont{J.}~\bibnamefont{Spalek}},
  \bibnamefont{and}
  \bibinfo{author}{\bibfnamefont{A.~M.}~\bibnamefont{Oles}},
  \bibinfo{journal}{J. Phys.: Solid State C} \textbf{\bibinfo{volume}{10}},
  \bibinfo{pages}{L271} (\bibinfo{year}{1977}).

\bibitem[{\citenamefont{Imada et~al.}(1989)\citenamefont{Imada}}]{Imada1989A}
  \bibinfo{author}{\bibfnamefont{M.}~\bibnamefont{Imada}},
  \bibnamefont{and}
  \bibinfo{author}{\bibfnamefont{Y.}~\bibnamefont{Hatsugai}},  
  \bibinfo{journal}{J. Phys. Soc. Jpn.} \textbf{\bibinfo{volume}{58}},
  \bibinfo{pages}{3752} (\bibinfo{year}{1989}).

\bibitem[{\citenamefont{Pryadko et~al.}(1989)\citenamefont{Pryadko}}]{Pryadko2004A}
  \bibinfo{author}{\bibfnamefont{L.}~\bibnamefont{Pryadko}},
  \bibinfo{author}{\bibfnamefont{S.}~\bibnamefont{Kivelson}},
  \bibnamefont{and}
  \bibinfo{author}{\bibfnamefont{O.}~\bibnamefont{Zachar}},  
  \bibinfo{journal}{Phys. Rev. Lett.} \textbf{\bibinfo{volume}{92}},
  \bibinfo{pages}{067002} (\bibinfo{year}{2004}).

\bibitem[{\citenamefont{Vedeneev et~al.}(1995)\citenamefont{Vedeneev}}]{Vedeneev1995A}
  \bibinfo{author}{\bibfnamefont{S.~I}~\bibnamefont{Vedeneev}},
  \bibinfo{author}{\bibfnamefont{A.~G.~M.}~\bibnamefont{Jansen}},
  \bibinfo{author}{\bibfnamefont{A.~A.}~\bibnamefont{Tsvetkov}},
  \bibnamefont{and}
  \bibinfo{author}{\bibfnamefont{P.}~\bibnamefont{Wyder}},  
  \bibinfo{journal}{Phys. Rev. B} \textbf{\bibinfo{volume}{51}},
  \bibinfo{pages}{16380} (\bibinfo{year}{1995}).

\bibitem[{\citenamefont{Hofer et~al.}(2000)\citenamefont{Hofer}}]{Hofer2000A}
  \bibinfo{author}{\bibfnamefont{J.}~\bibnamefont{Hofer}},
  \bibinfo{author}{\bibfnamefont{K.}~\bibnamefont{Conder}},
  \bibinfo{author}{\bibfnamefont{T.}~\bibnamefont{Sasagawa}},
  \bibinfo{author}{\bibfnamefont{G.}~\bibnamefont{Zhao}},
  \bibinfo{author}{\bibfnamefont{M.}~\bibnamefont{Willemin}},
  \bibinfo{author}{\bibfnamefont{H.}~\bibnamefont{Keller}},
  \bibnamefont{and}
  \bibinfo{author}{\bibfnamefont{K.}~\bibnamefont{Kishio}},  
  \bibinfo{journal}{Phys. Rev. Lett.} \textbf{\bibinfo{volume}{84}},
  \bibinfo{pages}{4192} (\bibinfo{year}{2000}).

\bibitem[{\citenamefont{Kulic}(2000)\citenamefont{Kulic}}]{Kulic2000A}
  \bibinfo{author}{\bibfnamefont{M.~L.}~\bibnamefont{Kulic}},
  \bibinfo{journal}{Phys. Rep.} \textbf{\bibinfo{volume}{338}},
  \bibinfo{pages}{1} (\bibinfo{year}{2000}).

\bibitem[{\citenamefont{Cuk et~al.}(2005)\citenamefont{Cuk}}]{Cuk2005A}
  \bibinfo{author}{\bibfnamefont{T.}~\bibnamefont{Cuk}},
  \bibinfo{author}{\bibfnamefont{D.~H.}~\bibnamefont{Lu}},
  \bibinfo{author}{\bibfnamefont{X.~J.}~\bibnamefont{Zhou}},
  \bibinfo{author}{\bibfnamefont{Z.~-X.}~\bibnamefont{Shen}},
  \bibinfo{author}{\bibfnamefont{T.~P.}~\bibnamefont{Deveraux}},
  \bibnamefont{and}
  \bibinfo{author}{\bibfnamefont{N.}~\bibnamefont{Nagaosa}},
  \bibinfo{journal}{Phys. Stat. Sol. (b)} \textbf{\bibinfo{volume}{242}},
  \bibinfo{pages}{11} (\bibinfo{year}{2005}).

%%%%%%%%%%%%%%%%%%%%%%%%%%%%%%%%%%%%%%%%%%%%%%%%%%%%%%%

%%%%%%%%%%%%%%%%%%%%%%%%%%%%%%%%%%%%%%%%%%%%%%%%%%%%%%%%%%%%%%%%%%%%%%%%%%%%%%%%%%%%%%%%%%%%%%%%%%%%%%%%%%%%%%%%%%%%%%%%%%%%%%%%%%%%%%%%%%%%%%%%%%%%%%%%%%%%%

\bibitem[{\citenamefont{Zhang et~al.}(1988)\citenamefont{Zhang, and Rice}}]{Zhang1988A}
  \bibinfo{author}{\bibfnamefont{F.~C.}~\bibnamefont{Zhang}},
  \bibnamefont{and}
  \bibinfo{author}{\bibfnamefont{T.~M.}~\bibnamefont{Rice}}, 
  \bibinfo{journal}{Phys. Rev. B} \textbf{\bibinfo{volume}{37}},
  \bibinfo{pages}{3759} (\bibinfo{year}{1988}).

\bibitem[{\citenamefont{Eskes et~al.}(1988)\citenamefont{Eskes, and Sawatzky}}]{Eskes1988A}
  \bibinfo{author}{\bibfnamefont{H.}~\bibnamefont{Eskes}},
  \bibnamefont{and}
  \bibinfo{author}{\bibfnamefont{G.~A.}~\bibnamefont{Sawatzky}}, 
  \bibinfo{journal}{Phys. Rev. Lett.} \textbf{\bibinfo{volume}{61}},
  \bibinfo{pages}{1415} (\bibinfo{year}{1988}).

\bibitem[{\citenamefont{Hybertsen et~al.}(1988)\citenamefont{Hybertsen, Schluter, and Christensen}}]{Hybertsen1988A}
  \bibinfo{author}{\bibfnamefont{M.~S.}~\bibnamefont{Hybertsen}},
  \bibinfo{author}{\bibfnamefont{M.}~\bibnamefont{Schluter}},
  \bibnamefont{and}
  \bibinfo{author}{\bibfnamefont{N.~E.}~\bibnamefont{Christensen}}, 
  \bibinfo{journal}{Phys. Rev. B} \textbf{\bibinfo{volume}{39}},
  \bibinfo{pages}{9028} (\bibinfo{year}{1989}).

\bibitem[{\citenamefont{Rice et~al.}(1988)\citenamefont{Hybertsen, Schluter, and Christensen}}]{Rice1991A}
  \bibinfo{author}{\bibfnamefont{T.~M.}~\bibnamefont{Rice}},
  \bibinfo{author}{\bibfnamefont{F.}~\bibnamefont{Mila}},
  \bibnamefont{and}
  \bibinfo{author}{\bibfnamefont{F.~C.}~\bibnamefont{Zhang}}, 
  \bibinfo{journal}{Philos. Trans. R. Soc. London, Ser. A} \textbf{\bibinfo{volume}{334}},
  \bibinfo{pages}{459} (\bibinfo{year}{1991}).

%%%%%%%%%%%%%%%%%%%%%%%%%%%%%%%%%%%%%%%%%%%%%%%%%%%%%%%%%%%%%%%%%%%%%%%%%%%%%%%%%%%%%%%%%%%%%%%%%%%%%%%%%%%%%%%%%%%%%%%%%%%%%%%%%%%%%%%%%%%%%%%%%%%%%%%%%%%%%

\bibitem[{\citenamefont{Newns et~al.}(1995)\citenamefont{Newns, Tsuei, and
  Pattnaik}}]{Newns}
\bibinfo{author}{\bibfnamefont{D.~M.} \bibnamefont{Newns}},
  \bibinfo{author}{\bibfnamefont{C.~C.} \bibnamefont{Tsuei}}, \bibnamefont{and}
  \bibinfo{author}{\bibfnamefont{P.~C.} \bibnamefont{Pattnaik}},
  \bibinfo{journal}{Phys. Rev. B} \textbf{\bibinfo{volume}{52}},
  \bibinfo{pages}{13611} (\bibinfo{year}{1995}).

\bibitem[{\citenamefont{Gasser et~al.}(1999)\citenamefont{Gasser, Heiner, and
  Elk}}]{Gasser1999A}
\bibinfo{author}{\bibfnamefont{W.}~\bibnamefont{Gasser}},
  \bibinfo{author}{\bibfnamefont{E.}~\bibnamefont{Heiner}}, \bibnamefont{and}
  \bibinfo{author}{\bibfnamefont{K.}~\bibnamefont{Elk}},
  \emph{\bibinfo{title}{Greensche Funktionen in Festk{\"o}rper- und
  Vielteilchenphysik}} (\bibinfo{publisher}{VILEY-VCH Verlag GmbH},
  \bibinfo{address}{Weinheim}, \bibinfo{year}{1999}).

\bibitem[{\citenamefont{Szcz{\c{e}}{\'s}niak and
  Durajski}(2014)}]{Szczesniak2012K}
\bibinfo{author}{\bibfnamefont{R.}~\bibnamefont{Szcz{\c{e}}{\'s}niak}}
  \bibnamefont{and} \bibinfo{author}{\bibfnamefont{A.~P.}
  \bibnamefont{Durajski}}, \bibinfo{journal}{Supercond. Sci. Technol.}
  \textbf{\bibinfo{volume}{27}}, \bibinfo{pages}{015003}
  (\bibinfo{year}{2014}).

\bibitem[{\citenamefont{Szcz{\c{e}}{\'s}niak
  et~al.}(2013)\citenamefont{Szcz{\c{e}}{\'s}niak, Durajski, and
  Szcz{\c{e}}{\'s}niak}}]{Szczesniak2012M}
\bibinfo{author}{\bibfnamefont{D.}~\bibnamefont{Szcz{\c{e}}{\'s}niak}},
  \bibinfo{author}{\bibfnamefont{A.~P.} \bibnamefont{Durajski}},
  \bibnamefont{and}
  \bibinfo{author}{\bibfnamefont{R.}~\bibnamefont{Szcz{\c{e}}{\'s}niak}},
  \bibinfo{journal}{J. Phys.: Condens. Matter} \textbf{\bibinfo{volume}{26}},
  \bibinfo{pages}{255701} (\bibinfo{year}{2014}).

\bibitem[{\citenamefont{Nakayama et~al.}(2009)\citenamefont{Nakayama, Sato,
  Terashima, Arakane, Takahashi, Kubota, Ono, Nishizaki, Takahashi, and
  Kobayashi}}]{Nakayama2009A}
\bibinfo{author}{\bibfnamefont{K.}~\bibnamefont{Nakayama}},
  \bibinfo{author}{\bibfnamefont{T.}~\bibnamefont{Sato}},
  \bibinfo{author}{\bibfnamefont{K.}~\bibnamefont{Terashima}},
  \bibinfo{author}{\bibfnamefont{T.}~\bibnamefont{Arakane}},
  \bibinfo{author}{\bibfnamefont{T.}~\bibnamefont{Takahashi}},
  \bibinfo{author}{\bibfnamefont{M.}~\bibnamefont{Kubota}},
  \bibinfo{author}{\bibfnamefont{K.}~\bibnamefont{Ono}},
  \bibinfo{author}{\bibfnamefont{T.}~\bibnamefont{Nishizaki}},
  \bibinfo{author}{\bibfnamefont{Y.}~\bibnamefont{Takahashi}},
  \bibnamefont{and}
  \bibinfo{author}{\bibfnamefont{N.}~\bibnamefont{Kobayashi}},
  \bibinfo{journal}{Phys. Rev. B} \textbf{\bibinfo{volume}{79}},
  \bibinfo{pages}{140503(R)} (\bibinfo{year}{2009}).

\bibitem[{\citenamefont{Nunner et~al.}(1999)\citenamefont{Nunner, Schmalian,
  and Bennemann}}]{Nunner1999A}
\bibinfo{author}{\bibfnamefont{T.~S.} \bibnamefont{Nunner}},
  \bibinfo{author}{\bibfnamefont{J.}~\bibnamefont{Schmalian}},
  \bibnamefont{and} \bibinfo{author}{\bibfnamefont{K.~H.}
  \bibnamefont{Bennemann}}, \bibinfo{journal}{Phys. Rev. B}
  \textbf{\bibinfo{volume}{59}}, \bibinfo{pages}{8859} (\bibinfo{year}{1999}).

\bibitem[{\citenamefont{Bohnen et~al.}(2003)\citenamefont{Bohnen, Heid, and
  Krauss}}]{Bohnen2003A}
\bibinfo{author}{\bibfnamefont{K.~P.} \bibnamefont{Bohnen}},
  \bibinfo{author}{\bibfnamefont{R.}~\bibnamefont{Heid}}, \bibnamefont{and}
  \bibinfo{author}{\bibfnamefont{M.}~\bibnamefont{Krauss}},
  \bibinfo{journal}{Europhys. Lett.} \textbf{\bibinfo{volume}{64}},
  \bibinfo{pages}{104} (\bibinfo{year}{2003}).

\end{thebibliography}
\end{document}